\documentclass[aps,twocolumn]{revtex4}
\usepackage{graphicx}
\begin{document}
\title{Neighborhood models of minority opinion spreading}
\author{C.J. Tessone\thanks{E-Mail: \tt tessonec@imedea.uib.es}$^{1}$, %
R. Toral \thanks{E-Mail: \tt raul@imedea.uib.es}$^{1,2}$, %
P. Amengual\thanks{E-Mail: \tt pau@imedea.uib.es}$^{1}$, %
H.S. Wio\thanks{Present address: Instituto de F\'{\i}sica de Cantabria,
Av. Los Castros s/n, E-39005 Santander, Spain; E-Mail: \tt
wio@ifca.unican.es}$^{1,2}$ and M. San Miguel\thanks{E-Mail: \tt maxi@imedea.uib.es}$^{1,2}$ %
}
%
%

\affiliation{$^1$Instituto Mediterr\'aneo de Estudios Avanzados IMEDEA
(CSIC-UIB), Campus UIB, E-07122 Palma de Mallorca, Spain\\
$^2$Departamento de F\'{\i}sica, Universitat de les Illes Balears, E-07122
Palma de Mallorca, Spain}
\date{\today}
%
\begin{abstract}We study the effect of finite size population in Galam's model [Eur. Phys. J. B \textbf{25} (2002) 403] of minority opinion spreading and introduce neighborhood models that account for local spatial effects. For
systems of different sizes $N$, the time to reach consensus is shown to scale
as $\ln N$ in the original version, while the evolution is much slower in the new
neighborhood models. The threshold value of the initial concentration of
minority supporters for the defeat of the initial majority, which is
independent of $N$ in Galam's model, goes to zero with growing system size in
the neighborhood models. This is a consequence of the existence of a critical
size for the growth of a local domain of minority supporters. \end{abstract}

\pacs{ {87.23.Ge}{Dynamics of social systems}, {0.5.50.+q} {Lattice theory and
statistics} }
%

\maketitle
\section{Introduction}

There is a growing interest among theoretical physicists in
complex phenomena in fields departing from the traditional realm
of physics research. In particular, the application of statistical
physics methods to social phenomena is discussed in several
reviews \cite{weidlich,Ball,stauffer04,Gal04}. One of the
sociological problems that attracts much attention is the building
or the lack of consensus out of some initial condition. There are
several different models that simulate and analyze the dynamics of
such processes in opinion formation, cultural dynamics, etc.
\cite{Gal03,Gal00,Gal01,Gal02,complex1,heg01,sznajd01,sznajd02,stauffer1,stauffer11,stauffer2,Castellano,Klemm1,Klemm2,Klemm3,krapi01,Mobilia1,Mobilia2}.
Among all those models, the one introduced by Galam
\cite{Gal01,Gal02} to describe the spreading of a minority
opinion, incorporates basic mechanisms of social inertia, resulting
in democratic rejection of social reforms initially favored by a
majority. In this model, individuals gather during their social
life in {\sl meeting cells} of different sizes where they discuss
about a topic until a final decision, in favor or against, is taken
by the entire group. The decision is based on the {\sl majority
rule} such that everybody in the meeting cell adopts the opinion
of the majority. Galam introduced the idea of ``social inertia" in
the form of a bias corresponding to a resistance to changes or
reforms, that is: in case of tie, one of the decisions (in the
original version, the one against) is systematically adopted. We
will describe in detail the model and its main conclusions in the
next sections. This simple model is able to explain why an
initially minority opinion can become a majority in the long run. An
interesting example was its application to the spread of rumors
concerning some September 11-th opinions in France \cite{Gal02}.
One of the major conclusions of the mean-field-like analysis in
Ref.\cite{Gal01}, is the existence of a threshold value $p_c <
1/2$ for the initial concentration of individuals with the
minority opinion (against the social reform). For $p > p_c$ every
individual eventually adopts the opinion of the initial minority,
so that the social reform is rejected and the {\sl status quo} is
maintained. A related message of this result is that a rumor
spreads, although initially supported by a minority, if the
society has some bias towards accepting it.

Galam traces back his results to dynamical effects produced by the
existence of asymmetric unstable points previously considered by
Granovetter \cite{Granovetter} and Schelling \cite{Schelling}.
These are fixed points of recursion relations describing the
dynamics of the fraction of a population adopting one of two
possible choices. In {\sl threshold models} these relations are
obtained considering a mean field type of interaction in which
individual thresholds to change choice (tolerance) are compared
with the fraction of the population that has already adopted the
new choice. Granovetter himself \cite{Granovetter} discusses that
the stability of the fixed points can be changed by spatial
effects, noting that the assumption that each individual is
responsive to the behavior of all the others is often
inappropriate. In \cite{Gal01} such complete connectedness of the
population seems to be circumvented by the introduction of the
meeting cells. Only individuals in each meeting cell interact
among themselves. In this sense each meeting cell plays the role
of a {\sl bounded neighborhood} \cite{Schelling} and it is still
possible to obtain analytically recursion relations for the
dynamics. However, contrary to the {\sl bounded neighborhood}
model of Schelling, individuals enter and leave these
neighborhoods randomly, and the neighborhoods do not have any
characteristic identity other than their sizes. Even if the
meeting cells are thought of as sites where local discussions take
place, Galam's model \cite{Gal01} does not incorporate local
interactions since the individuals are randomly redistributed in
the meeting cells at each time step of the dynamics.

The alternative considered by Schelling to the {\sl bounded
neighborhood} model is a {\sl spatial proximity} model in which
everybody defines his neighborhood by reference to his own spatial
location. The spatial arrangement or configuration within the
neighborhood mediates the interactions. We propose here a
different neighborhood model which shares some characteristics
with the {\sl bounded neighborhood} and {\sl spatial proximity}
models: The meeting cells are neighborhoods defined by spatial
location, therefore introducing important local effects,  but the
interaction within the neighborhood is independent of the spatial
configuration within the cell. Contrary to the model in
\cite{Gal01} the individuals are here located at fixed sites of a
lattice. The local neighborhood or meeting cell in which a given
individual interacts changes with time, reflecting neighborhoods
of changing shape and size. Such neighborhood model could be
appropriate for a relatively primitive society in which
interactions are predominantly among neighbors, but the size of
the neighborhood or interaction range is not fixed.

A different version of Galam's model was introduced by Stauffer
\cite{stauffer11}. At variance with our neighborhood models in which
individuals are fixed in the sites of a lattice and the meeting cells have a
maximum size, Stauffer considers the situation in which individuals freely
diffuse in a lattice with only a fraction of sites being occupied. This
diffusion process leads to the formation of ``natural" clusters which play the
role of our meeting cells:  It is within each one of these clusters in which
the rule of majority opinion and bias towards minority in case of a tie are
taken. Stauffer finds in his model that the time to reach a consensus opinion
grows logarithmically with system size $N$. We also find this dependence in the
original model of Galam, while in our neighborhood models the consensus time
takes much larger values and is compatible with a power law dependence. A
related model, including the figure of the ``contrarians" (that is, people that
always oppose to the majority position), was later introduced by Galam
\cite{Gal04b} and also analyzed by Stauffer \cite{stauffer4}.

A main consequence of introducing the spatial effects considered
in our neighborhood models is that the threshold found in
\cite{Gal01} disappears with system size, i.e. $\lim_{N \to \infty} p_c
=0$, so that in large systems the minority opinion always spreads
and overcomes the initial majority for whatever initial proportion
of the minority opinion. This is a consequence of the existence of
a critical size for a local domain of minority supporters. Domains
of size larger than the critical one will expand and occupy the
whole system. For large systems there is always a finite
probability to have a domain of over-critical size in the initial
condition. While in traditional Statistical Physics we are mostly
concerned with the thermodynamic limit of large systems, these
findings emphasize the important role of system size in the
sociological context of models of interacting individual entities.

The outline of the paper is as follows. Section 2 reviews the
original definition of Galam's model \cite{Gal01,Gal02} and
introduces our new local neighborhood models. In Sect. 3 we go
beyond the mean field limit of Refs.\cite{Gal01,Gal02} by
discussing the system size dependence of the predictions of the
original model. Steady-state and dynamical properties of our
neighborhood models are presented in Sects. 4 and 5. General
conclusions are summarized in Sect. 6.

\section{Definition of the model}
\subsection{Galam's original non-local model}

The model considers a population of $N$ individuals who randomly
gather in ``meeting cells". A meeting cell is just defined by the
number of individuals $k$ that can meet in the cell. Let us define
$a_k$ as the probability that a particular person is found in a
cell of size $k$. Obviously, it is $\sum_k a_k=1$.

The dynamics of the model is as follows: first the meeting cells
are defined by giving each one a size according to the probability
distribution $\{a_k\}$, such that the sum of all the cell sizes
equals the number of individuals $N$, but otherwise their location
or shape are not specified. These cells are not modified during
the whole dynamical process. The persons have an initial binary
(against, $+$, or in favor, $-$) opinion about a certain topic.
The probability that a person shares the $+$ opinion at time $t$
is $P_+(t)$ and an equivalent definition for $P_-(t)=1-P_+(t)$.
Initially one sets $P_+(t=0)=p$. Alternatively, $P_+(t)$ can be
thought of as the proportion of people supporting opinion $+$ at
time $t$.

The $N$ individuals are then distributed randomly among the
different cells. The basic premise of the model is that all the
people within a cell adopt the opinion of the majority of the
cell. Furthermore, in the case of a tie (which can only occur if
the cell size $k$ is an even number), one of the opinions, that we
arbitrarily identify with the $+$ opinion, is adopted. Once an
opinion within the different cells has been taken, time increases
by one, $t\to t+1$ and the individuals rearrange by distributing
themselves again randomly among the different cells.

The main finding of this model is that an initially minority
opinion, corresponding to $p < 1/2$ can win in the long term. This
is an effect of the tie rule that selects the $+$ opinion in case
of a tie.

It is possible to write down a recursion relation for the density
of people that at time $t$ have the $+$ opinion as \cite{Gal01}
\begin{equation}\label{map:mas}
P_+(t+1) = \sum_{k=1}^M a_k \, \sum_{j=\left[\frac{M}{2}+1\right]}^M
{k\choose j} P_+(t)^j \, \left[1-P_+(t)\right]^{k-j},
\end{equation}
Simultaneously\footnote{These expressions correct a misprint in
Eqs. (4,5) of reference \cite{Gal01}.}
\begin{equation}\label{map:menos}
P_-(t+1) = \sum_{k=1}^{M} a_k \,
\sum_{j=\left[\frac{M+1}{2}\right]}^M{k\choose j} P_-(t)^j \,
\left[1-P_-(t)\right]^{k-j},
\end{equation}
the notation $[x]$ indicates the integer part of $x$. This is a
mean-field equation that neglects possible fluctuations.

For a wide range of distributions $\{a_k\}$ this map has three
fixed points: two stable ones at $P_+=1$ and $P_+=0$ and an
unstable one, the {\em faith point}, at $P_+= p_c$. Hence, the
dynamics is such that
\begin{equation} \label{eq:nl:infty}
\lim_{t \to \infty} P_+(t) = %
\left\{ %
\begin{array}{lr} %
1 & \hbox{if } \qquad P_+(0)>p_c\\
0 & \hbox{if } \qquad P_+(0)<p_c
\end{array} %
\right.
\end{equation}


\subsection{Neighborhood models}\label{neigh-models}

\begin{figure}
\begin{center}
\includegraphics[width=7cm]{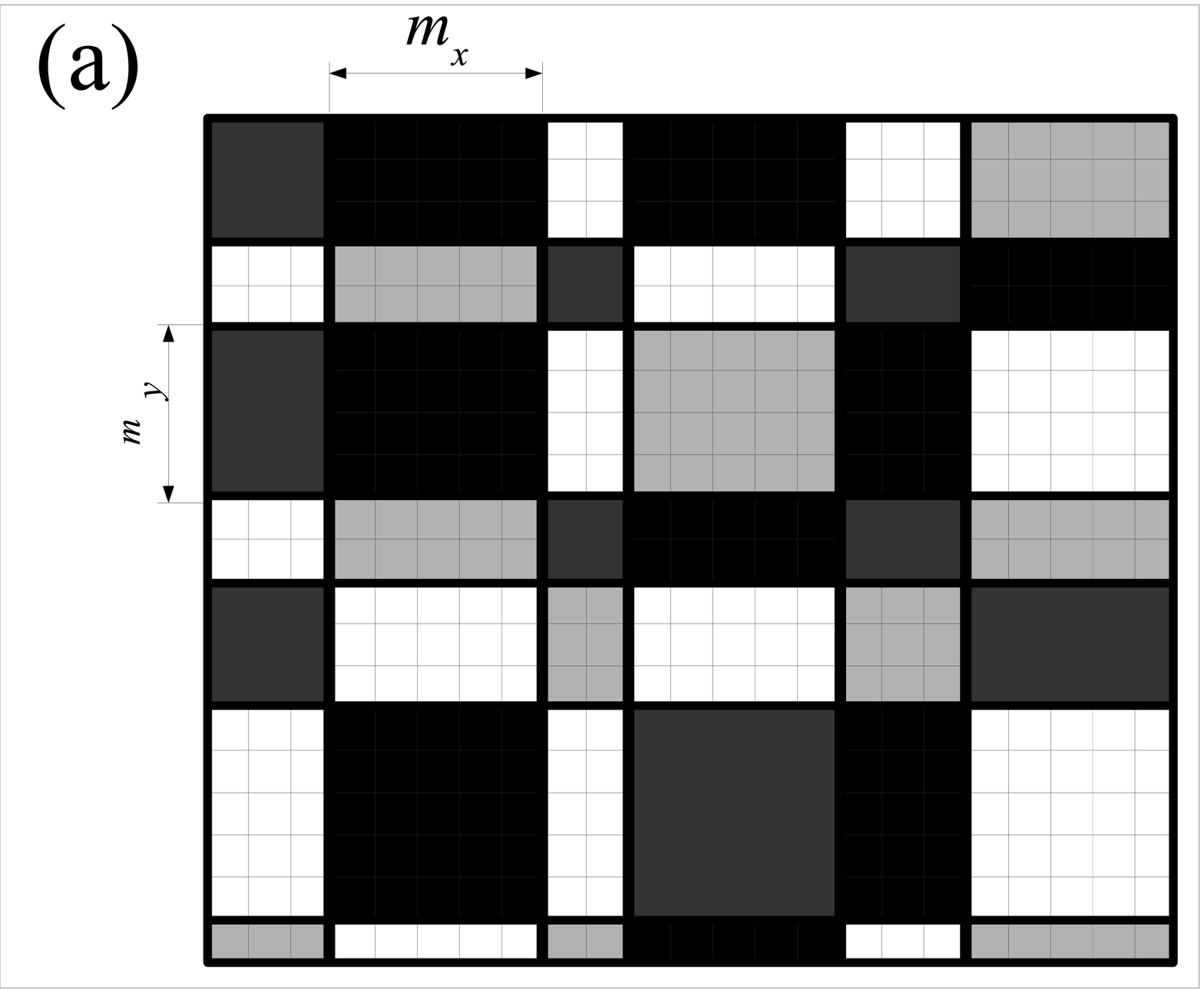}\\
\includegraphics[width=7cm]{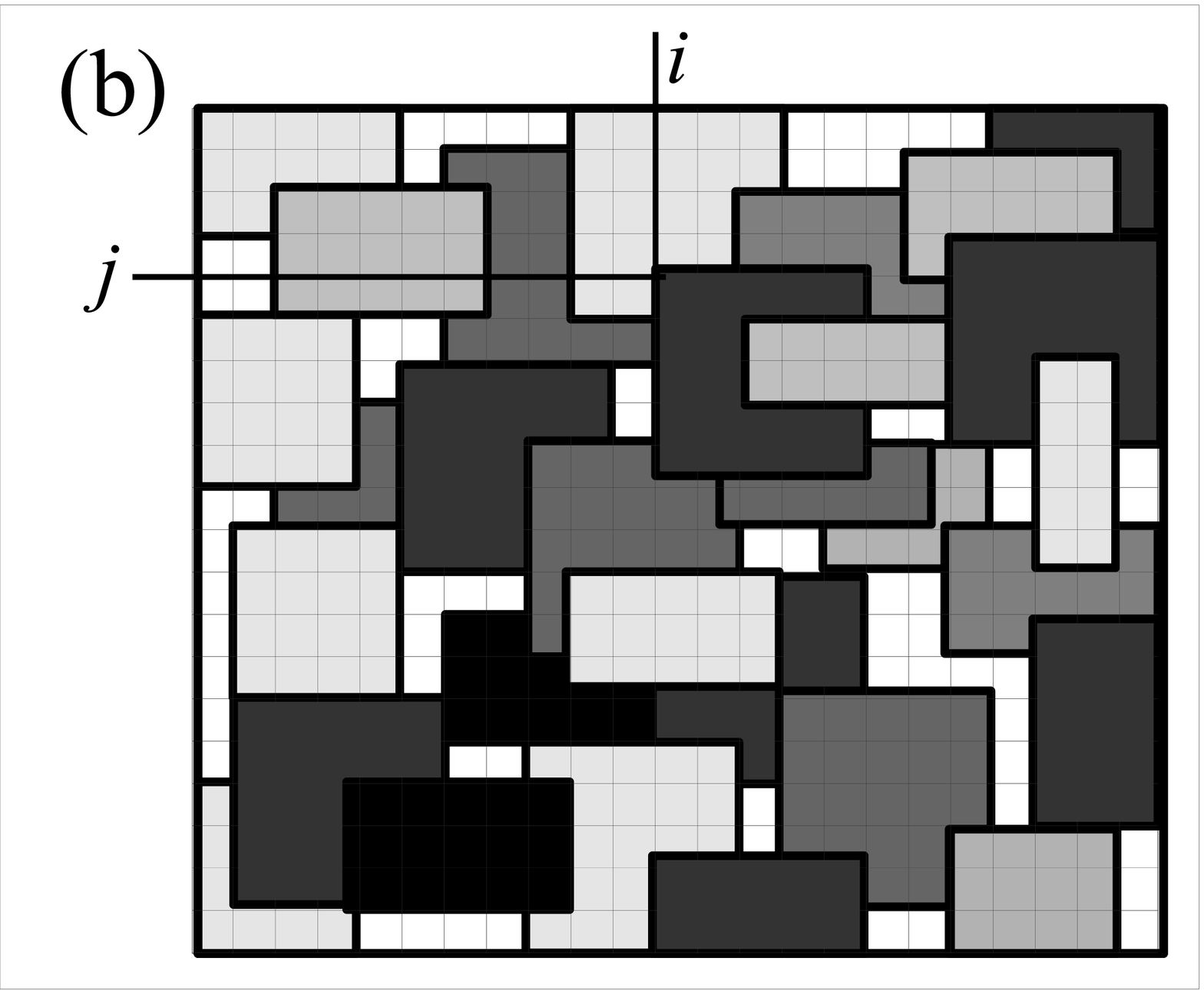}
\end{center}
\caption{\label{tes-fig}(a) Regular 2D tessellation: All the cells
are simultaneously created, being $m_x$ and $m_y$ uniformly
distributed between 1 and $M$. (b) Locally grown tessellation: A
site ($i,j$) is chosen, and from it a cell of size $m_x,m_y$,
excluding those already belonging to other cell.}
\end{figure}

We now introduce our Neighborhood Models that incorporate local
spatial effects in the interacting dynamics proposed by Galam. In
these local models, individuals are fixed at the sites of a
regular lattice and they interact with other individuals in their
spatial neighborhood. We have considered several cases:

\subsubsection{One-dimensional neighborhood model: synchronous
update}

The $N$ individuals are distributed at the sites of a linear
lattice $i=1,2,\dots,N$. Once distributed, they never move again.
Initially they are assigned a probability $p$ of adopting the $+$
opinion, and $(1-p)$ the $-$ opinion. The dynamics starts by
defining the meeting cells $k=1,2,\dots$ as the segments
$[i_{k},i_{k+1}-1]$ of length $m_k=i_{k+1}-i_k$. The cell sizes
$m_k$ are distributed according to a uniform distribution in the
interval $[1,M]$. The average cell size is hence $\langle m_k
\rangle=\frac{(M+1)}{2}$ and the average number of cells is
$N/\langle m_k \rangle$. Once the cells are defined, the dynamical
rules of Galam's model are applied synchronously to all the cells,
time increases by one $t\to t+1$. In the next time step, new
cells, uncorrelated to the previous ones are defined and the
dynamical rules applied again. The process continues until there
is a single common opinion in the whole system.

\subsubsection{Two-dimensional neighborhood model: synchronous
update}

The two-dimensional case is very similar to the 1D version
explained above. The only difference is the way the meeting cells
are defined in the two-dimensional lattice. An individual is now
characterized by two indexes $(i,j)$ with $1\le i,j \le L$, such
that the total number of individuals is $N=L^2$. We have
considered two different definitions of the cells originated in
two tessellations of the plane: (a) the {\sl regular}
tessellation and (b) the {\sl locally-grown} tessellation. In the
regular tessellation, we define segments in the $i$ and $j$ axis
independently, such that the sizes are in both cases uniformly
distributed between $1$ and $M$ in the same way that we did in the
one-dimensional case. Figure \ref{tes-fig}a plots a typical
example. In the {\sl locally-grown tessellation} we first choose a
site of the lattice and then define a rectangle around it whose
sides are both uniformly distributed between $1$ and $M$. The cell
is defined then as the sites in the resulting rectangle excluding
those sites that already were part of a previously defined cell.
Figure \ref{tes-fig}b shows a typical example.

Once the cells are defined, the dynamical rules are applied synchronously to
all the cells and a common opinion is formed within each cell. Time then
increases by one $t\to t+1$. In the next time step, new cells are defined and
the process continues until a consensus opinion is reached in the whole
population.

\subsubsection{Asynchronous update}

The 1D and 2D models have been also considered in the asynchronous update
version. In this case, a lattice site is randomly chosen and a cell defined
around it as a segment (1D) of size $m$ or a rectangle (2D) of size
$(m_x,m_y)$. It is only within this cell that the biased majority rule is
applied. Time increases by $t\to t+m/N$ in 1D and by $t\to t+(m_x \, m_y)/N$ in
2D. Then a new site is selected randomly and the process iterates until a
consensus opinion is obtained.

\section{Results for Galam's original non-local model.}

We present in this section an analysis of Galam's original
non-local model. Our aim is to go beyond the mean-field approach
of references \cite{Gal01,Gal02,Gal04b} by studying the system
size dependence of the different magnitudes of interest. Some of
the results are based on numerical simulations of the model.

\begin{figure}
\begin{center}
\includegraphics[width=7cm,angle=-90]{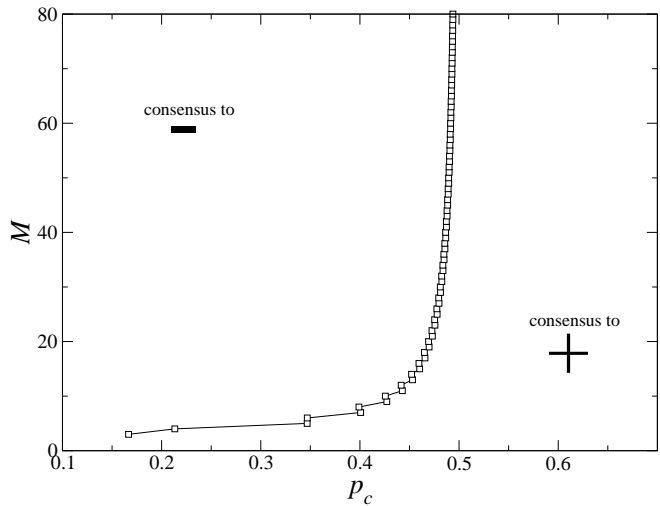}
\end{center}
\caption{\label{nl-M-pc} Phase diagram of the original
Galam's model in the plane ($p_c$, $M$). }
\end{figure}

We consider $N$ individuals that distribute themselves randomly in
meeting cells whose size is {\em uniformly distributed} between
$1$ and $M$. In the notation of Eq. (1), this means that $a_k= 2k
/(M(M+1))$, $k \in [1,M]$, as it follows from the fact that $a_k$
measures the probability for an individual of being in any of the
$k$ sites of a cell of size $k$. Initially we assign to each of
the persons any of the two possible opinions, such that the
probability of having the favored opinion is $p$. Again, in the
language of Eq. (\ref{map:mas}), we are setting $P_+(0)=p$. We
then apply the dynamical rules of Galam's model until a consensus
opinion is formed. By iteration of this procedure, we measure the
probability $\rho$ that the consensus opinion coincides with the
favored one, $+$. This is precisely defined as the fraction of
realizations that end up in the favored $+$ opinion.

\begin{figure}
\begin{center}
\includegraphics[width=7cm,angle=-90]{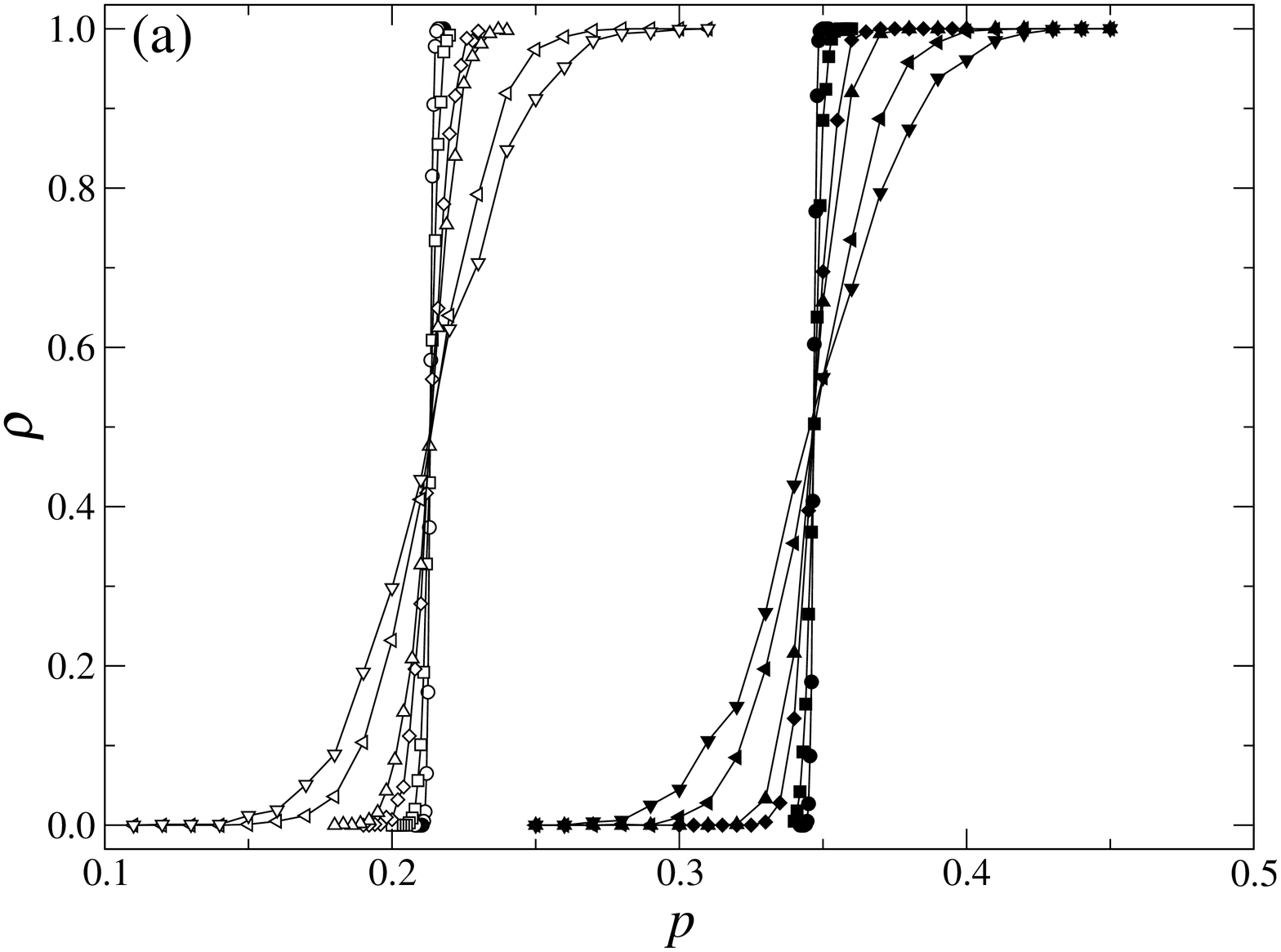}\\
\includegraphics[width=7cm,angle=-90]{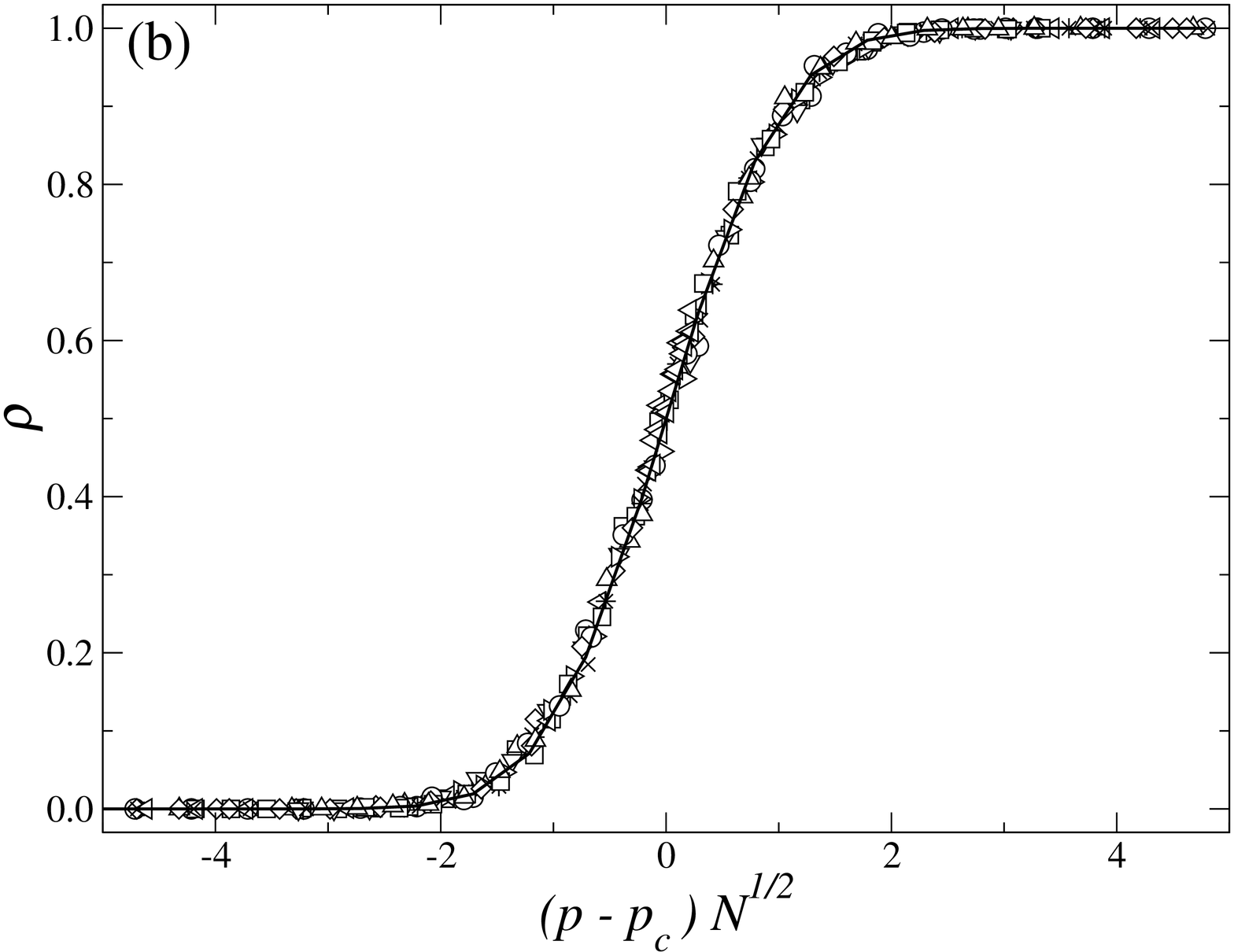}
\end{center}
\caption{\label{opnl}(a) Order parameter in Galam's non-local
model. The white symbols correspond to the case $M=4$
($p_c=0.2133077908\dots$), while the black ones to $M=5$
($p_c=0.3467871056\dots$). The values of $N$ range between
$N=10^3$ and $N=10^6$. (b) The order parameter for both values of
$M$ for rescaled values of $p$ according to $(p-p_c) N^{1/2}$.
Also, it is shown a fit with the function $\rho(p,N) =
(1+\hbox{erf}(x/1.17))/2 $ with the scaling variable
$x=(p-p_c)/N^{1/2}$. }
\end{figure}

The analysis of Eq. (\ref{map:mas}) predicts a first order phase transition in
the sense that the ``order parameter'' $\rho=0$ if $p < p_c$ and $\rho=1$ if $p
> p_c$. In Fig. \ref{nl-M-pc} we show, in the parameter space $(p_c,M)$, the
regions where the two solutions, as obtained by finding numerically the
non-trivial fixed point of the recurrence Eq. (\ref{map:mas}), exists. Note
that, as expected, the larger the decision cells, the closer the faith point to
$1/2$. Notice also in this figure that some pairs of consecutive values of $M$
give almost the same value for $p_c$. The reason is that, for odd values of
$M$, the rule that applies in case of a tie is used only up to the value $M-1$
(which is even), so odd numbers give similar values of $p_c$ than the precedent
even number.

Since Eq. (\ref{map:mas}) neglects possible system size fluctuations, Eq.
(\ref{eq:nl:infty}) this result is only valid in the limit $N\to \infty$. We
plot in Fig. \ref{opnl}a the raw results of our simulations for different
system sizes. The analysis carried out in Fig. \ref{opnl}b shows that the
asymptotic results of $N \to \infty$ are achieved by means of the following
scaling relation \begin{equation} \rho(p,N) = f\left( (p-p_c) \, N ^{-1/2}
\right). \end{equation} Therefore, there is a region of size $N^{-1/2}$ where
there is a significant probability that the results differ from the infinite
size limit. 

\begin{figure}
\begin{center}
\includegraphics[width=7cm,angle=-90]{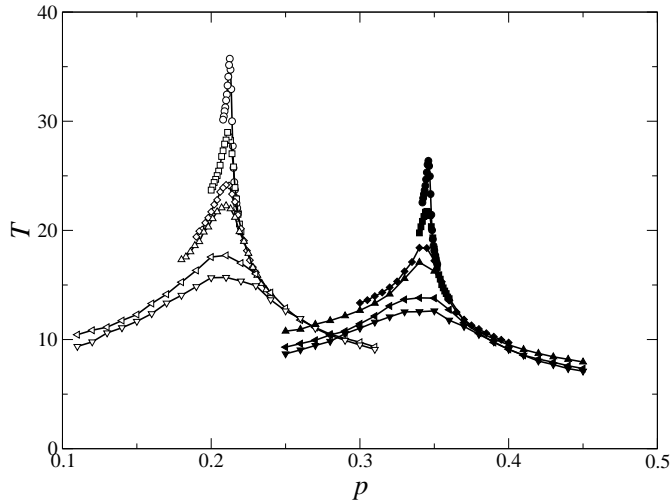}
\caption{\label{timenl} Time that it takes to the system to reach
the consensus state as a function of the initial probability $p$
for different system sizes. The white symbols correspond to the
case $M=4$ while the black ones to $M=5$. The values of $N$ range
between $N=10^3$ and $N=10^6$.}
\end{center}
\end{figure}

\begin{figure} \begin{center}
\includegraphics[width=7cm,angle=-90]{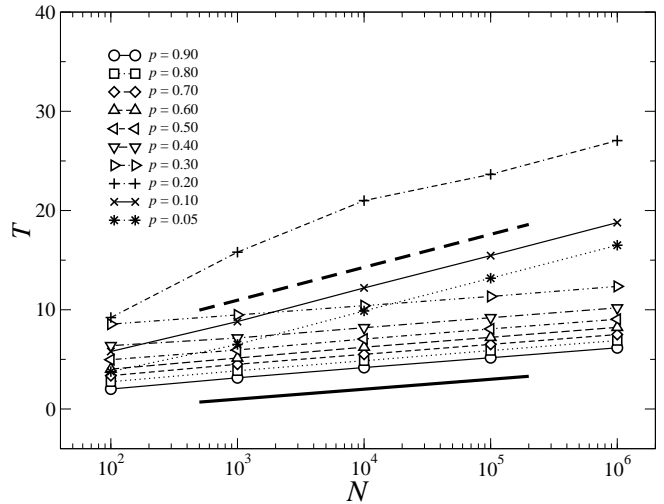}
\caption{\label{slopes} Time to reach the consensus state as a function of
system size for different values of $p$ and $M=4$. Two different slopes can be
seen for values of $p$ greater and lower than the critical value $p_c \sim
0.21$. Also the predicted slopes are plotted.} \end{center} \end{figure}

We now analyze the time $T$ it takes to reach the consensus
opinion. Strictly speaking, in the mean field approach the number
of iterations needed for Eq. (1) to converge to the fixed point is
infinity. In Ref. \cite{Gal01} it was adopted the criterion that
the fixed point had been reached at the time $T$ such that
$P_+(T)$ differs from the fixed point in less that $0.01$. In our
simulations, and as in reference \cite{stauffer11}, it is natural
to define the time $T$ as the finite number of steps needed to
reach the consensus opinion. We plot in Fig. \ref{timenl} the time
$T$ as a function of the initial probability $p$ of the favored
opinion. It can be seen that the time $T$ increases with
increasing system size and takes its maximum value at the faith
point. A closer analysis shown in figure \ref{slopes} shows that,
for all values of $p$, the time needed to reach the consensus
increases logarithmically with $N$. It is interesting to note that
the same logarithmic dependence was found in Ref.
\cite{stauffer11} for the model accounting for Brownian diffusion.

A simple argument can help to understand this logarithmic
behavior. One can mimic the size dependence of the time $T$ by
noticing that the definition used in the numerical simulations is
equivalent to define $T$ as the value for which $P_+(T)={\mathcal
O}(1/N)$ or $1-P_+(T)={\mathcal O}(1/N)$, since $1/N$ is the
minimum possible value for $p$ over $p=0$. This can be now
obtained by linearizing the evolution equation around any of the
fixed points $0,1$. Defining respectively $\delta(t) = P_+(t) - 0$
or $\delta(t) = P_+(t) - 1$, and replacing in the evolution
equation (\ref{map:mas}) $P_+(t+1)\equiv F[P_+(t)]$, we obtain at first order:
\begin{equation}
\delta(t+1) = \lambda \,\delta(t),
\end{equation}
where $\lambda\equiv F'[0]$ and $\lambda\equiv F'[1]$,
respectively. In this linear regime, we have simply: $\delta(t) =
\lambda^{t} \, \delta(0)$ and according to the definition above,
\begin{equation}
T \sim \frac {\ln \delta(0) - \ln N}{\ln \lambda},
\end{equation}
where the correct value of $\lambda$ has to be used in each case.
This is the logarithmic behavior observed in the simulations.
Furthermore, the slope of the logarithmic law will depend on the
fixed point at which the system tends. So, at one side of the
critical point, the slope should be different than at the other.
Figure \ref{slopes} shows the confirmation of this prediction,
where it is seen that for high values of $p$ (above the faith
point) the slope is quite different than for lower values. For
$M=4$, the predicted values for $\lambda$ are $\lambda=1/10$ and
$\lambda=1/2$ for the fixed point at $P_+=1$ and $P_+=0$
respectively. The corresponding slopes, $-1/\ln \lambda$, are
$0.434\dots$ and $1.442\dots$. As shown in the figure, these
values agree well with the measured slopes. The only discrepancy
is for $p\approx p_c$ for which the time needed to reach consensus
must include as well the time needed to leave the fixed point
$p_c$.

\section{Neighborhood models: steady state properties}

In this section we consider the steady state properties of our
neighborhood models defined in Sect. \ref{neigh-models}. We will
see that the introduction of spatial local effects leads to a very
different behavior than for the non-local version of the previous
section in which individuals were distributed randomly in fixed
cells.

\begin{figure}
\begin{center}
\includegraphics[width=7cm,angle=-90]{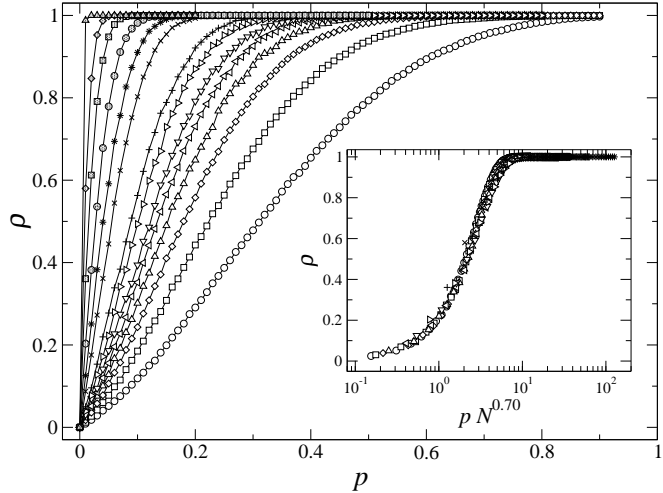}
\caption{\label{op1d} Order parameter for the neighborhood
one-dimensional system with synchronous update for $M=5$. The
system size ranges between $N=10$ (rightmost curve) and $N=10^4$
(leftmost curve). The inset shows the validity of the scaling law
$\rho=\rho(pN^{\alpha})$ with $\alpha=0.7$.}
\end{center}
\end{figure}

\begin{figure}
\begin{center}
\includegraphics[width=6.8cm,angle=-90]{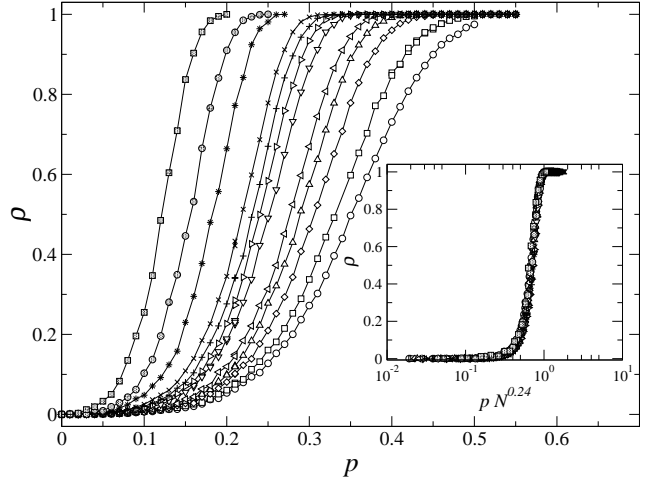}
\end{center}
\caption{\label{op2d} Same as Fig. \ref{op1d} for neighborhood
models with synchronous update and regular 2D tessellation. $M=5$.
The system size is $N=L\times L$ with $L$ between $L=15$
(rightmost curve) and $L=10^3$ (leftmost curve). The inset shows
the scaling law with $\alpha=0.24$.}
\end{figure}

\begin{figure}
\begin{center}
\includegraphics[width=7cm,angle=-90]{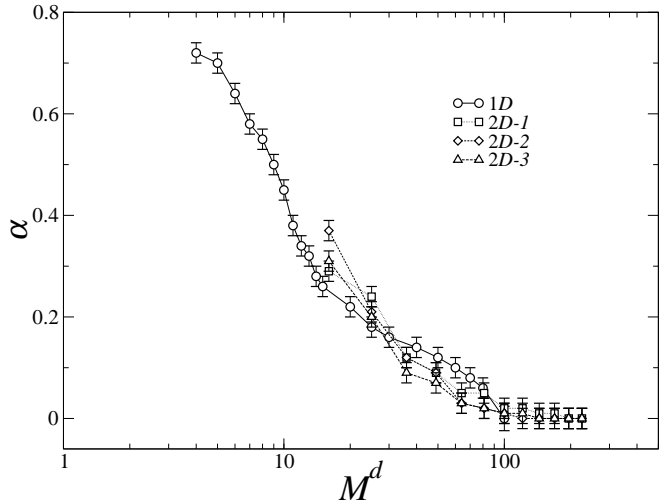}
\end{center}
\caption{\label{alpha} Scaling exponent defined in equation (\ref{eq:rho:p}) as a function of $M^d$. The different curves correspond to: 1D (one-dimensional system, synchronous update), while the bi-dimensional are: 2D-1 (synchronous update, regular
tessellation), 2D-2 (synchronous update, locally grown
tessellation), 2D-3 (asynchronous update). }
\end{figure}

In our neighborhood versions, similarly to the original Galam
model, it turns out that a consensus opinion is always reached in
a finite number of steps. We first consider the order parameter,
$\rho$, defined as the probability that the consensus opinion
coincides with the favored one. Figures \ref{op1d} and \ref{op2d}
show that, both in the 1D and the 2D cases, the order parameter
$\rho$ is an increasing function of the initial probability $p$ of
adopting the favored opinion. It is possible to define, quite
arbitrarily, the transition point $p_c$ as the one for which
$\rho=1/2$. However, $p_c$ depends upon the system size as a power
law $p_c(N)\sim N^{-\alpha}$, hence for increasing $N$ the
transition point tends to $p_c=0$. In other words, the transition
disappears in the thermodynamic limit, $N\to\infty$, and the
favored opinion, in that limit, is always the selected one
independently of the initial choice. More precisely, the data can
be described by the scaling law
\begin{equation}\label{eq:rho:p}
\rho(p,N)=\rho(p N^{\alpha}).
\end{equation}
The exponent $\alpha$ depends on the dimension and on the maximum
size $M$ of the domains. In Fig. \ref{alpha} we plot the
$M$-dependence of $\alpha$ in several 1D and 2D neighborhood models.
Notice that $\alpha$ decreases for increasing $M$ and it depends
on the system dimension, but it is basically independent of the
local rules defined. Alternatively, for fixed $p$ we can define a critical value $N_c(p)$ such that a small population $N<N_c(p)$ tends not to propagate the initially minority opinion.

\begin{figure} \begin{center}
\includegraphics[width=6.8cm,angle=-90]{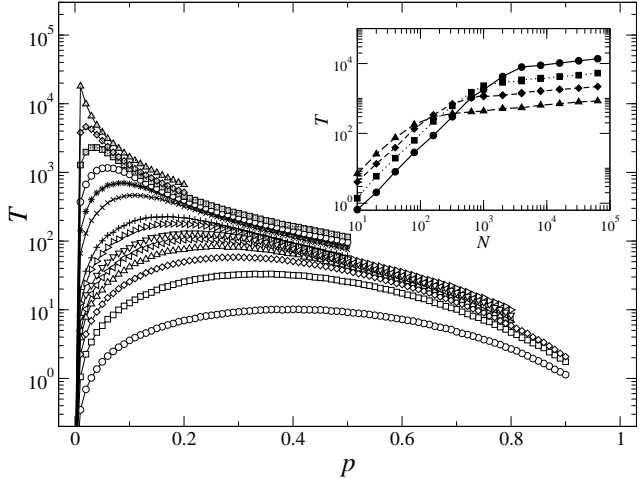}  \end{center}
\caption{\label{time1d} Time to reach the consensus vs $p$, for synchronous 1D
local cells and for $M=5$. Same symbols and systems sizes than in Fig.
\ref{op1d}.  The inset shows the time to reach consensus plotted against system
size, for different values of $p$, i.e. $p=0.2$ (triangles),  $p=0.1$
(diamonds),  $p=0.05$ (circles),  $p=0.02$ (squares). Two power-laws are
observed (depending on whether $N<N_c(p)$ or $N>N_c(p)$); for $N<N_c$,  $\beta
\approx 1.6$ and in the regime $N>N_c(p)$, $\beta \approx 0.2$. }   \end{figure}

\begin{figure}
\begin{center}
\includegraphics[width=6.8cm,angle=-90]{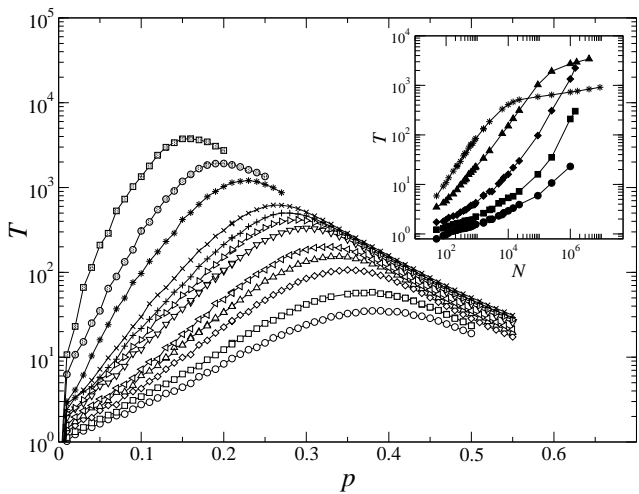}
\end{center}
\caption{\label{time2d} Time to reach the final consensus state vs
$p$, for the $2D$ cells, and for $M=5$. Same symbols and systems
sizes than in Fig. \ref{op2d}. The inset shows the time to reach consensus for fixed values of $p$: $p=0.3$ (stars),  $p=0.2$ (triangles),  $p=0.1$ (diamonds),  $p=0.05$ (circles),  $p=0.02$ (squares). The results are quite similar to those depicted in Figure \ref{time1d}. In this case, for $N<N_c(p)$, $\beta \approx 0.6$ and in the regime $N>N_c(p)$ (a regime only clearly seen in the cases $p=0.3,0.2$), $\beta \approx 0.1$. }
\end{figure}

\section{Neighborhood models: dynamical evolution}

There are several differences between the evolution in the neighborhood and
non-local models. 
We first analyze the time $T$ needed to reach the consensus. As in Galam's
original model, the data in Figs. \ref{time1d} and \ref{time2d} for the 1D and
2D cases, respectively, show that for fixed $N$, the time $T$ reaches a maximum
at the critical point $p_c(N)$. Notice that the numerical values for $T$ are
much larger in the local models that they were in the original model.
Furthermore, as shown in the insets of Figs. \ref{time1d} and \ref{time2d}, for
fixed $p$, the time $T$ has two different growth laws according to whether the
population $N$ is smaller or larger than the critical size $N_c(p)$. For
$N<N_c(p)$ the time $T$ increases as a power-law $T\sim N^{\beta}$ of the
system size $N$ with $\beta=1.6(0.6)$ for $d=1(2)$. For $N>N_c(p)$
the data are compatible with a power law with smaller exponents,
although it can not be completely excluded a logarithmic dependence in this
case. 

\begin{figure}
\begin{center}
 \includegraphics[width=2.2cm]{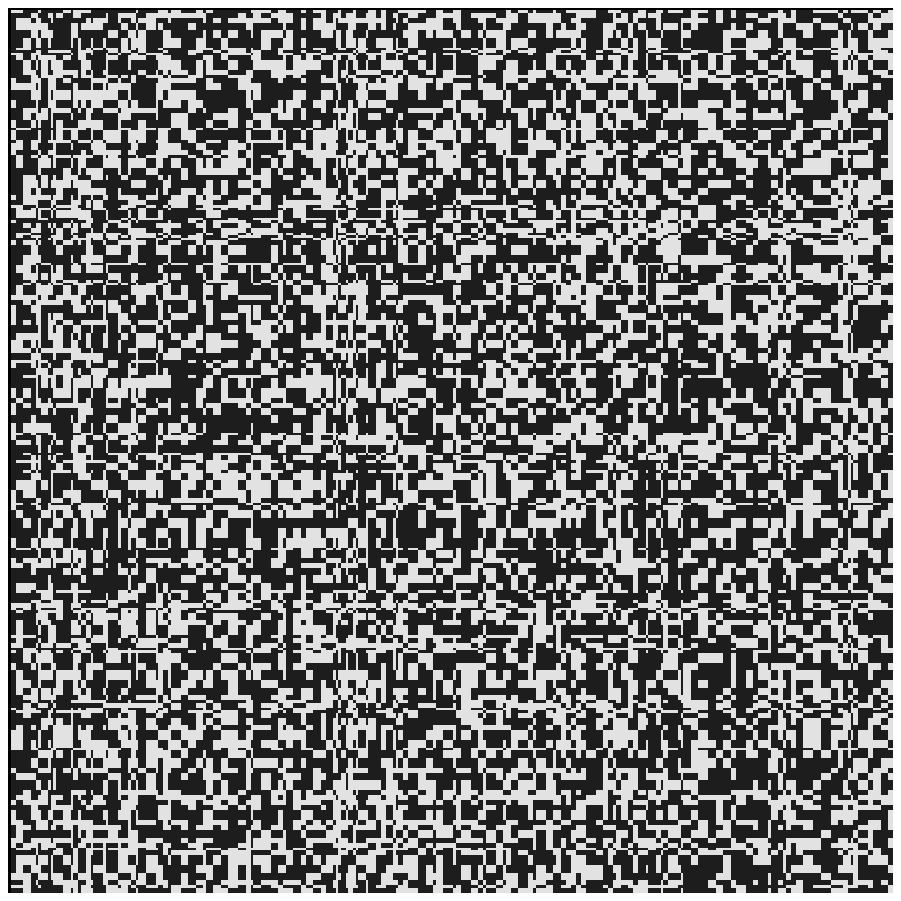}
 \includegraphics[width=2.2cm]{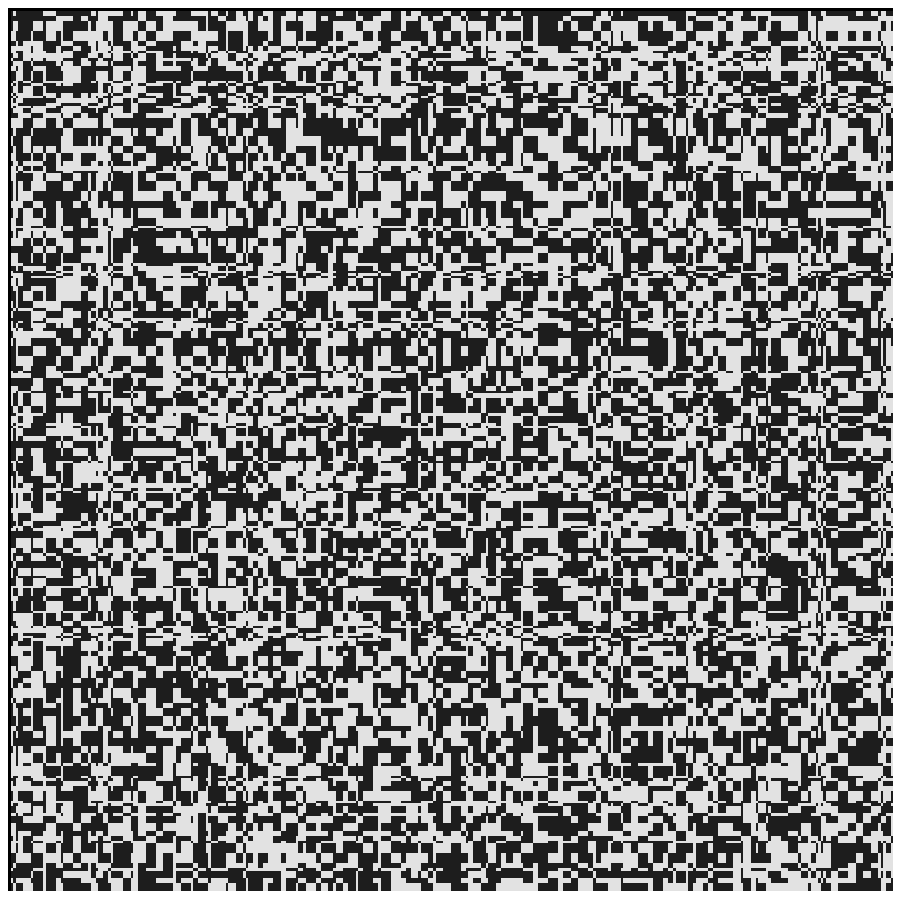}
 \includegraphics[width=2.2cm]{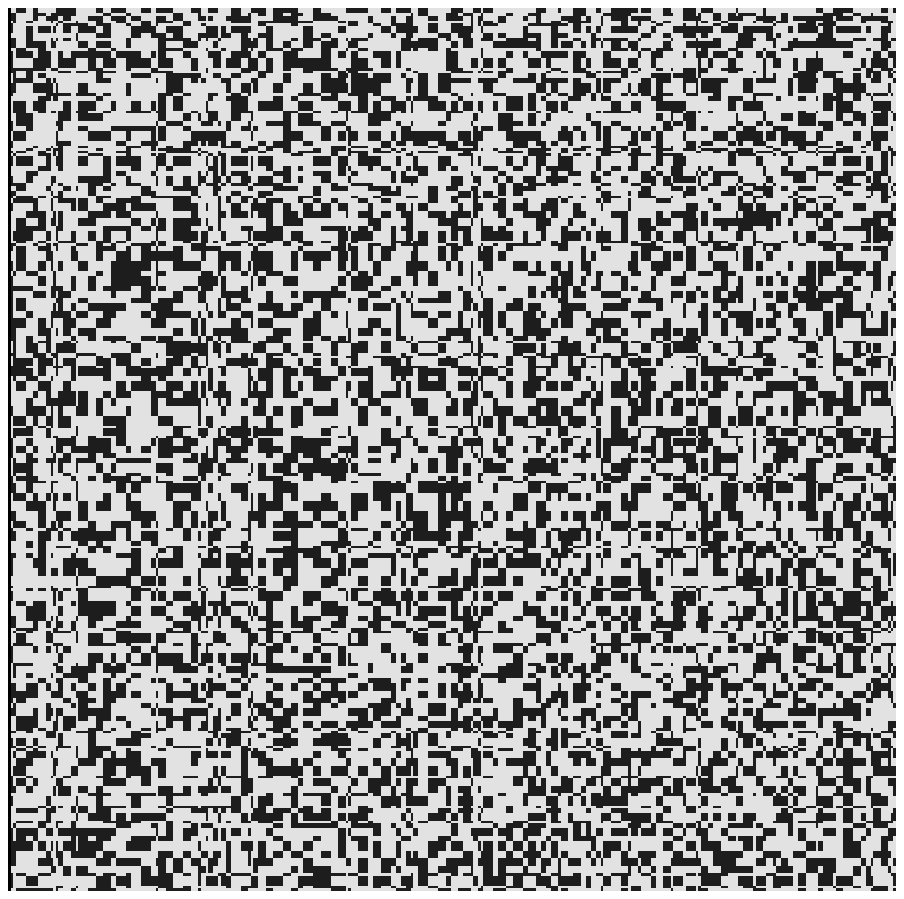}\\
 \includegraphics[width=2.2cm]{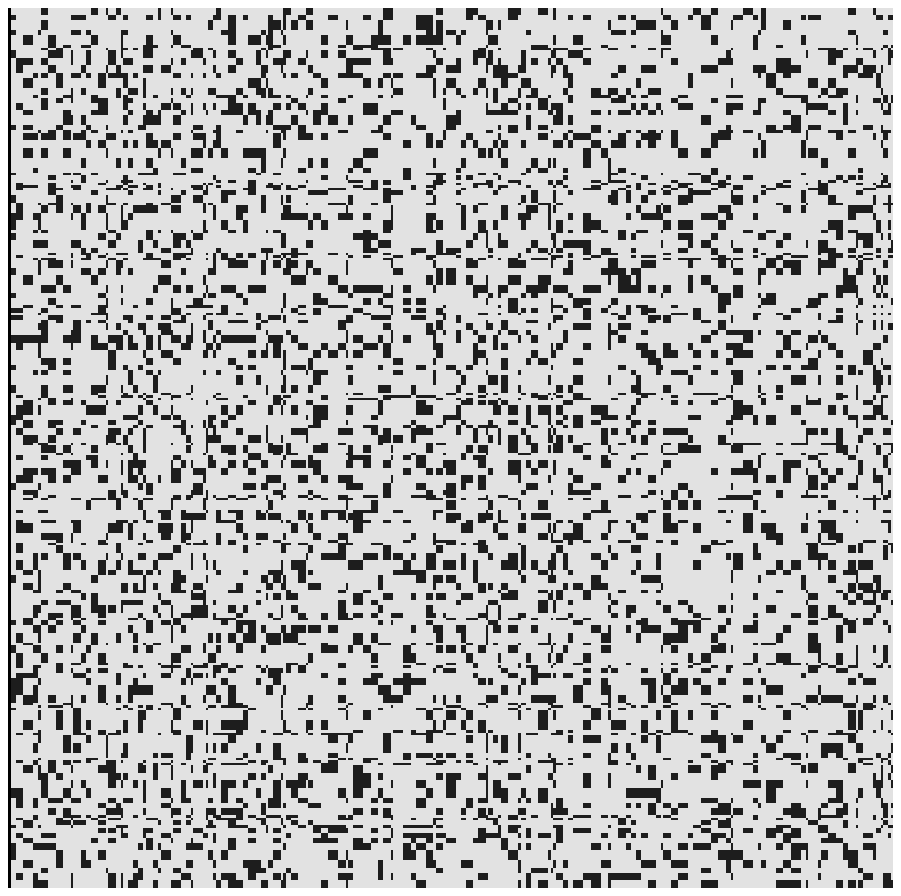}
 \includegraphics[width=2.2cm]{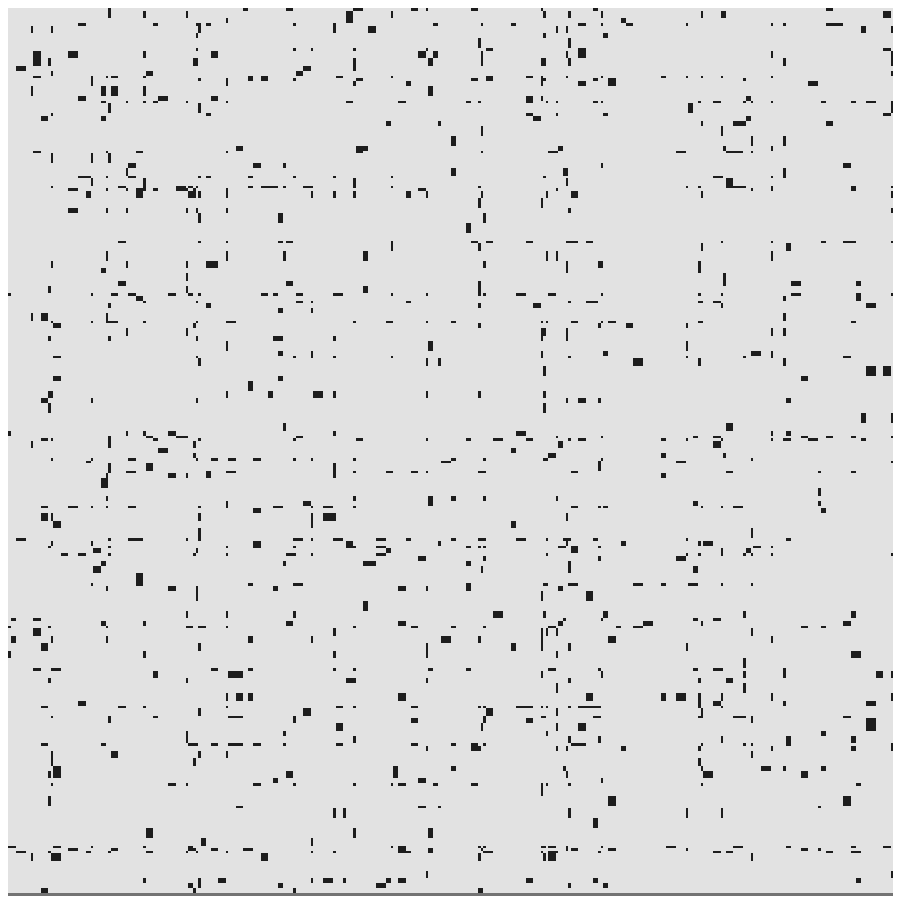}
 \includegraphics[width=2.2cm]{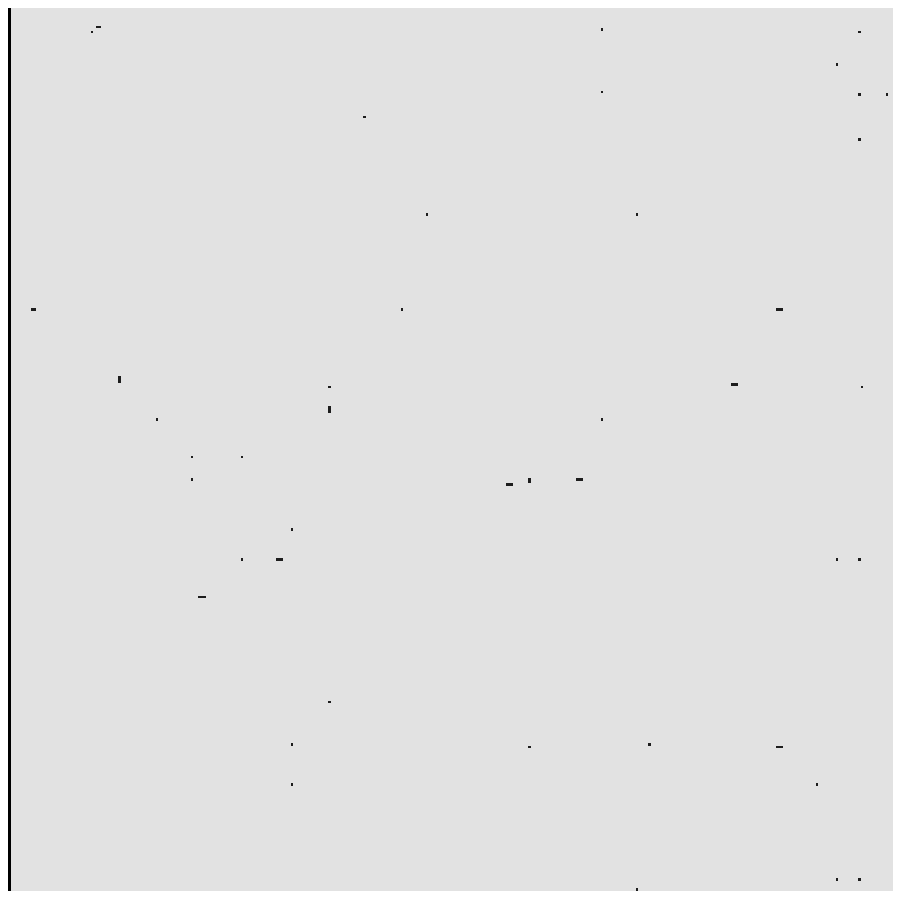}
\end{center}
\caption{\label{snap-nl} Plot of the dynamical evolution of a
non-local system updated according to the synchronous update.
Each frame correspond to a time, increasing from left to right and
top to bottom. The system size is $N=256^2$ and $M=5$.}
\end{figure}

In Fig. \ref{snap-nl} we plot several snapshots
corresponding to the evolution according to Galam's original rules. It is seen
that the evolution is very fast and at each time step the number of people
favoring a particular opinion increases but due to the non-locality of the
rules, one can not see any structured pattern of growth.

When using the neighborhood rules, however, it can be seen that,
after a very short transient in which domains are formed, the
ulterior evolution is by modification of the interfaces between
the two possible types of domains.

\begin{figure}
\begin{center}
 \includegraphics[width=6.8cm]{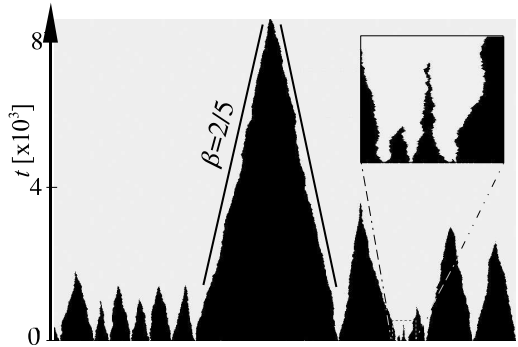}
\end{center}
\caption{\label{snap-1d} Plot of the dynamical evolution of a
local 1D system. Each row corresponds to a time, increasing from
bottom to top. The system size is $N=5000$ and $M=5$. It is also shown the predicted slope given by Eq. (\ref{eq:beta})
The inset amplifies a small region for a short time evolution.}
\end{figure}

Figure \ref{snap-1d}  depicts the dynamical evolution of the
synchronous 1D version of the neighborhood model. It shows a
linear growth of the size of domains of favored opinion. It is easy to
predict the slope of the linear growth according to the following
reasoning: the only evolution is produced if the discussion cell
intercepts two domains with different opinions. Let us consider
the position of an interface at time $t=0$. For cells of odd size,
the bias rule does not apply and the interface does not move in
the average. For cells of even size, the bias rule acts only in
one case and it is easy to show that the location of the
interface, on the average, increases by $1/2$. Therefore, the
size of the domain of the favored opinion increase as $t/2$. Averaging for an
equiprobability of having cells of size between 1 an $M$, yields a
linear growth $\beta t$ with
\begin{equation}\label{eq:beta}
\beta = 2\frac{\left[\frac{M}{2}\right]
\left[\frac{M}{2}+1\right]}{M(M+1)},
\end{equation}
where $[x]$ once again denotes the integer part of $x$. The validity of this result for $M=5$ is shown in Fig. \ref{snap-1d}.

\begin{figure}
\begin{center}
 \includegraphics[width=2.2cm]{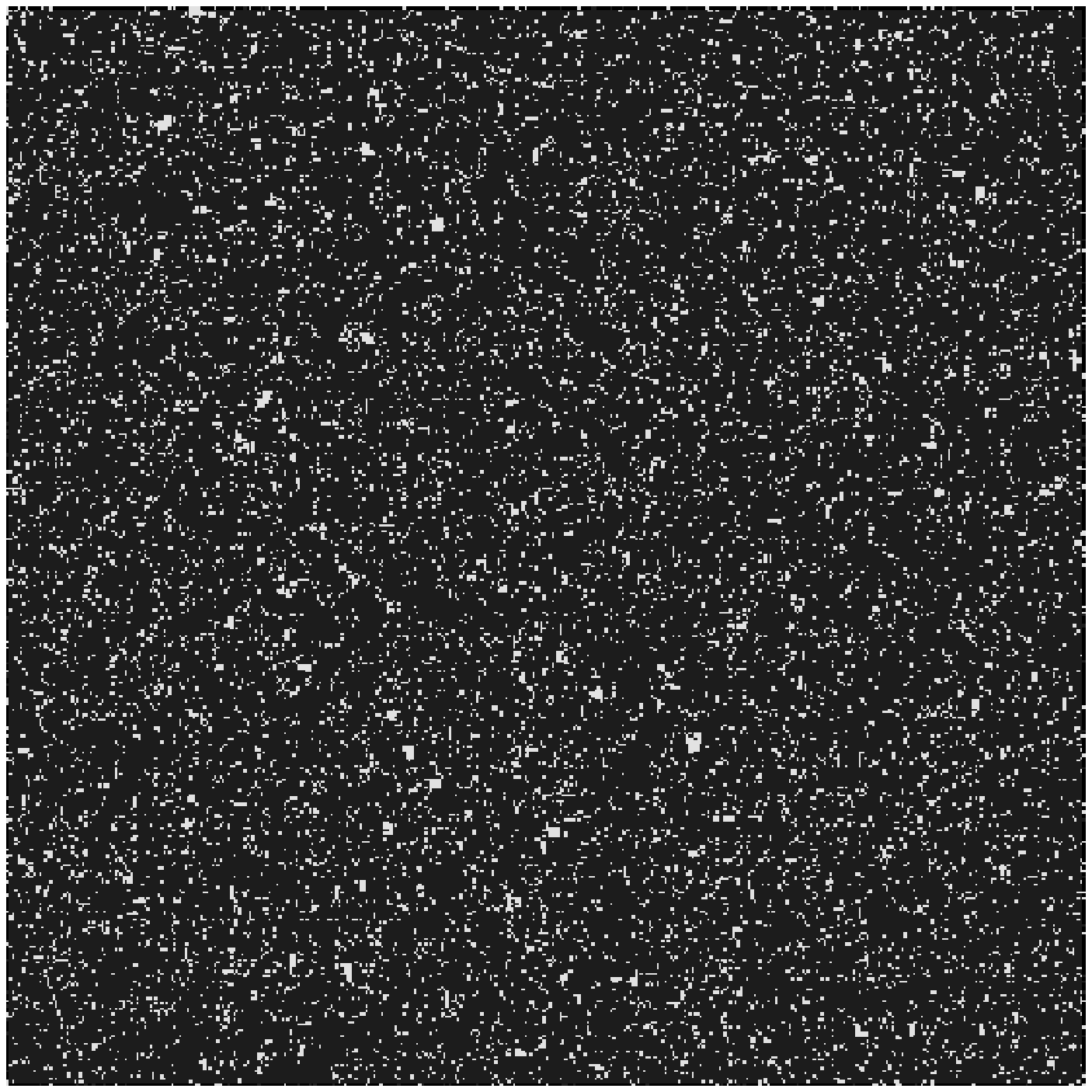}
 \includegraphics[width=2.2cm]{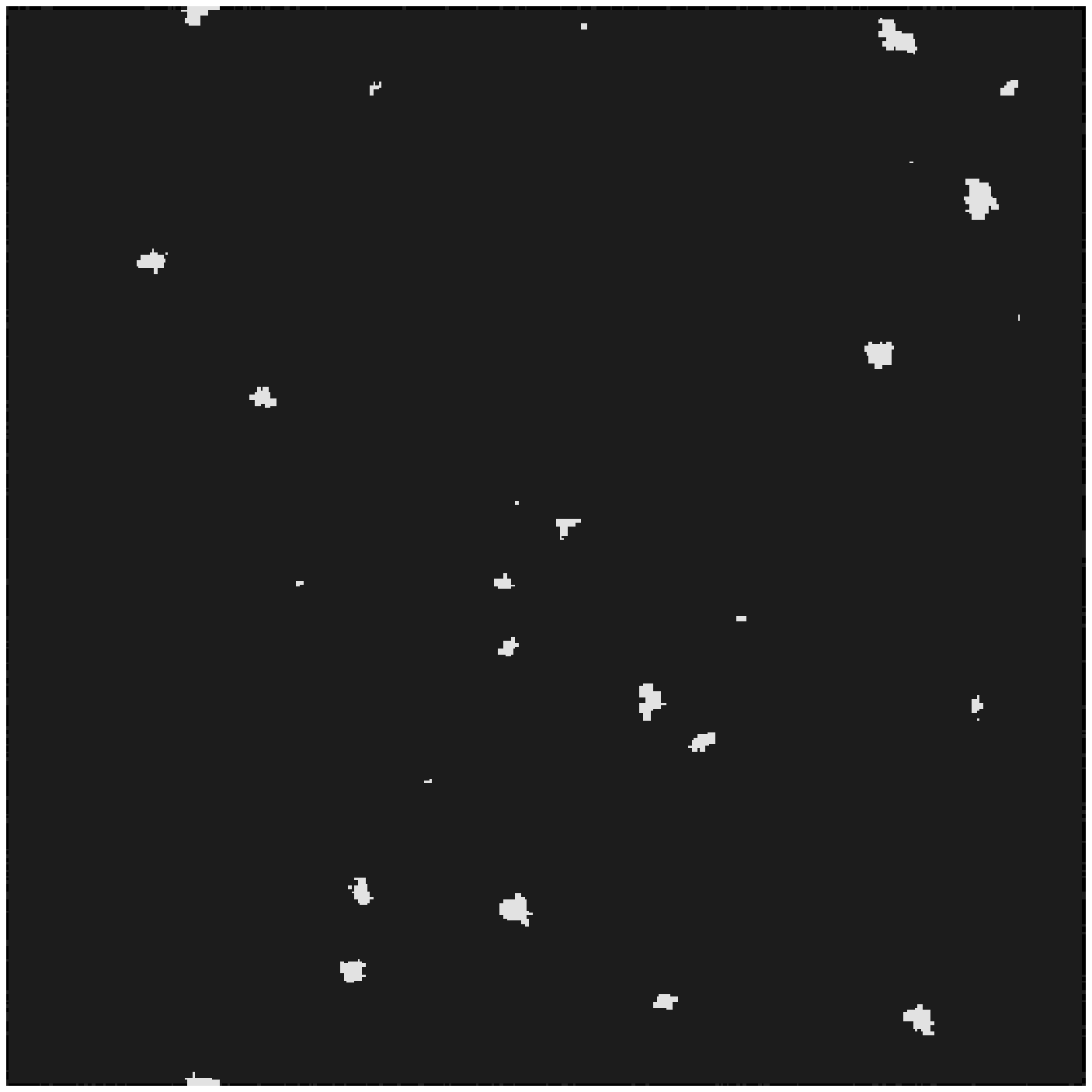}
 \includegraphics[width=2.2cm]{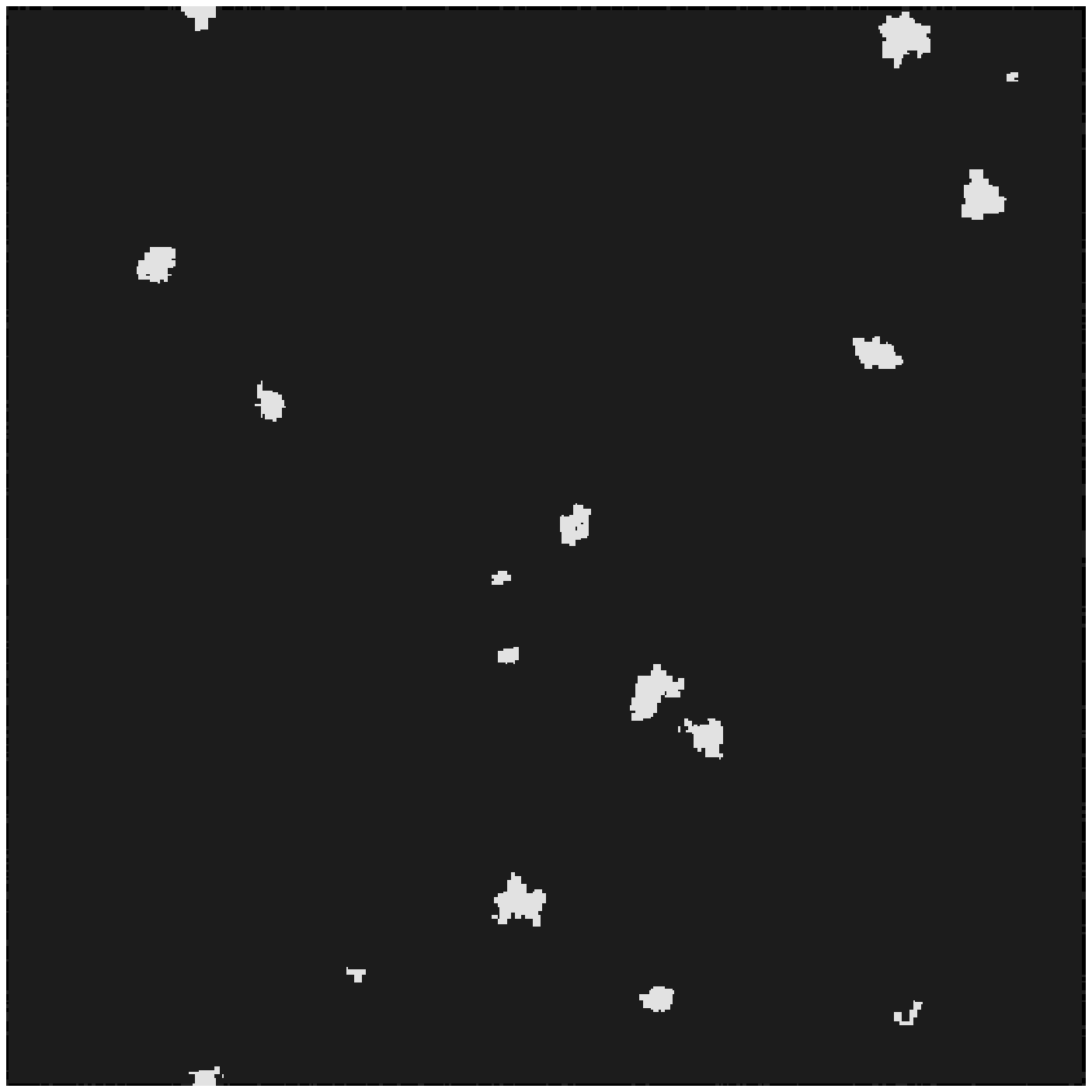}\\
 \includegraphics[width=2.2cm]{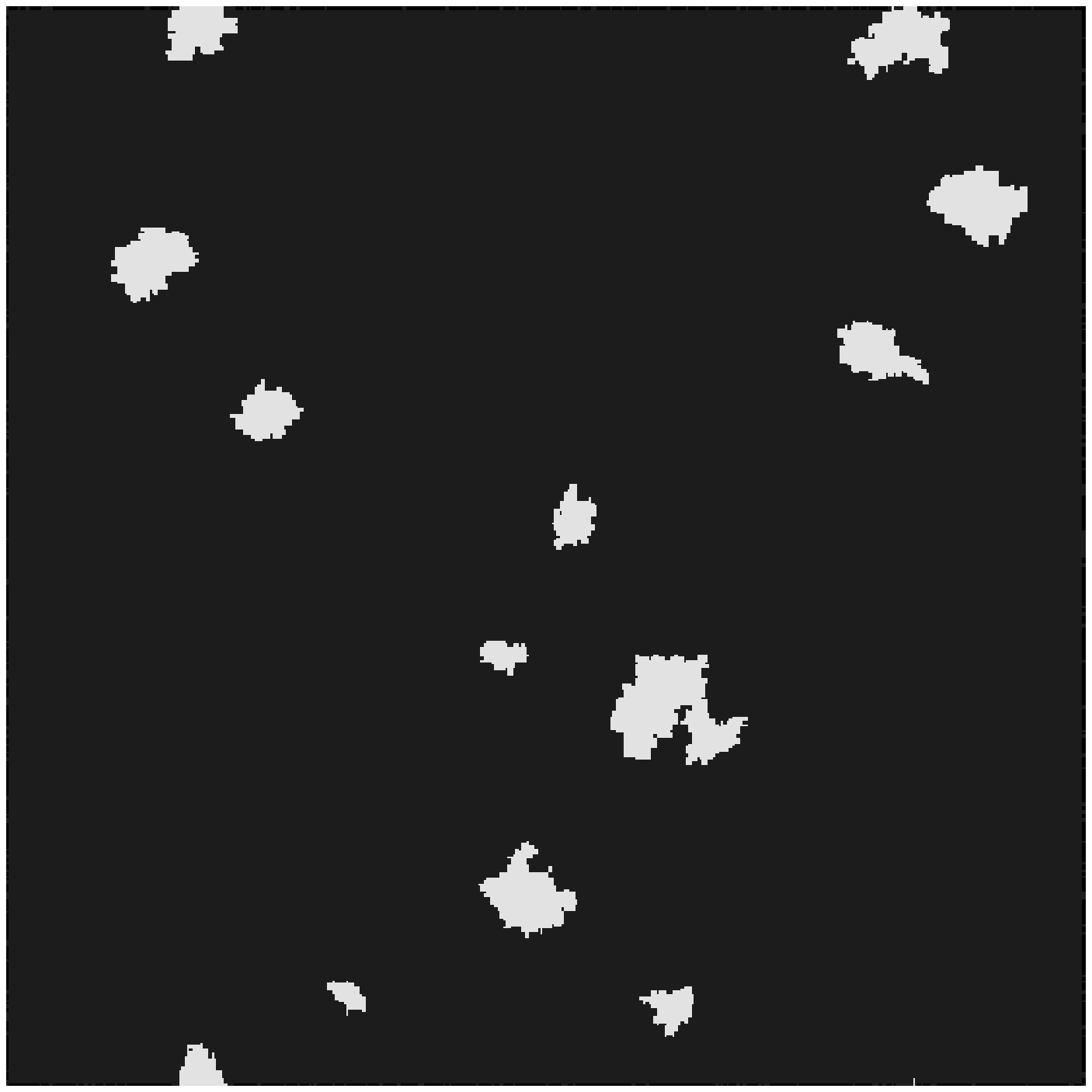}
 \includegraphics[width=2.2cm]{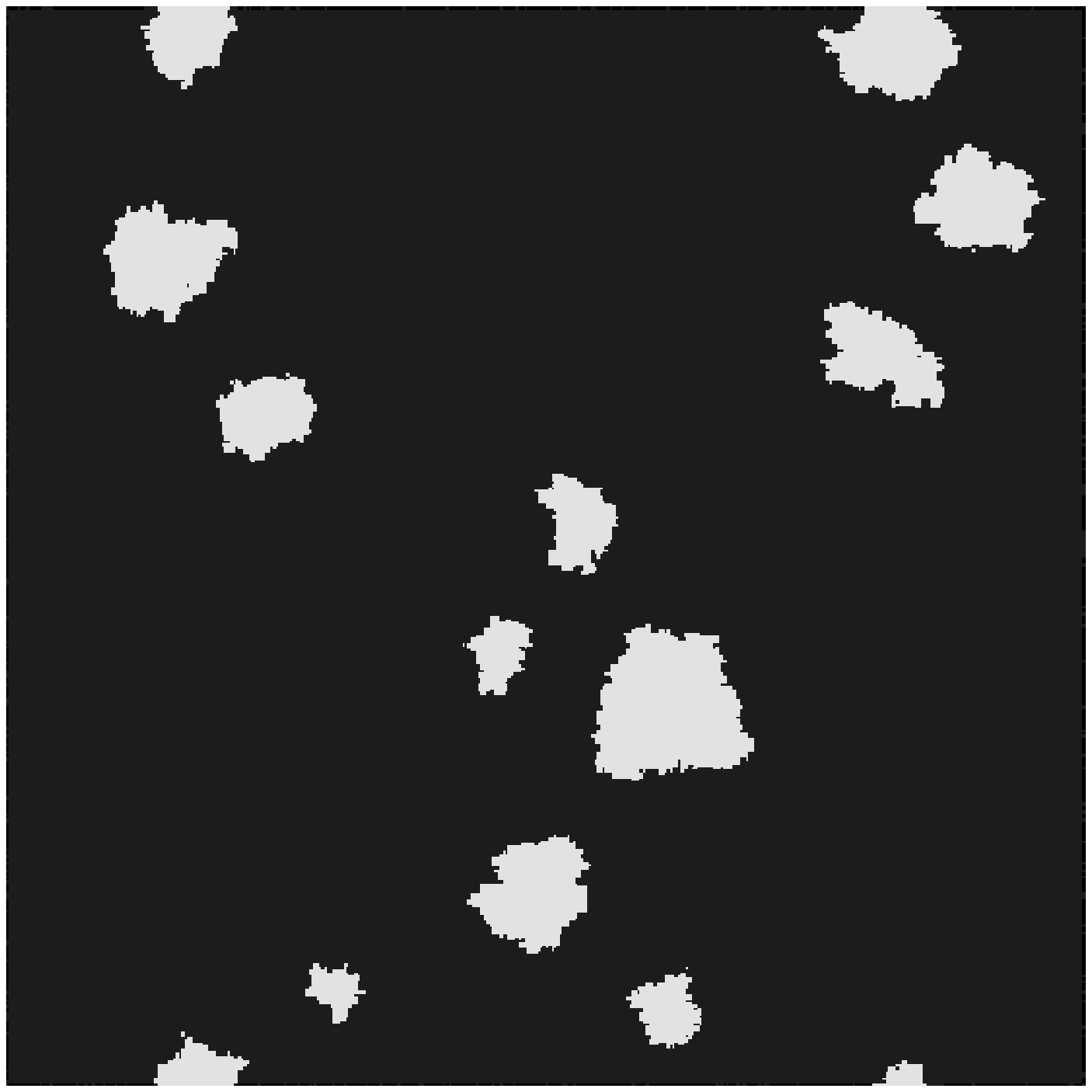}
 \includegraphics[width=2.2cm]{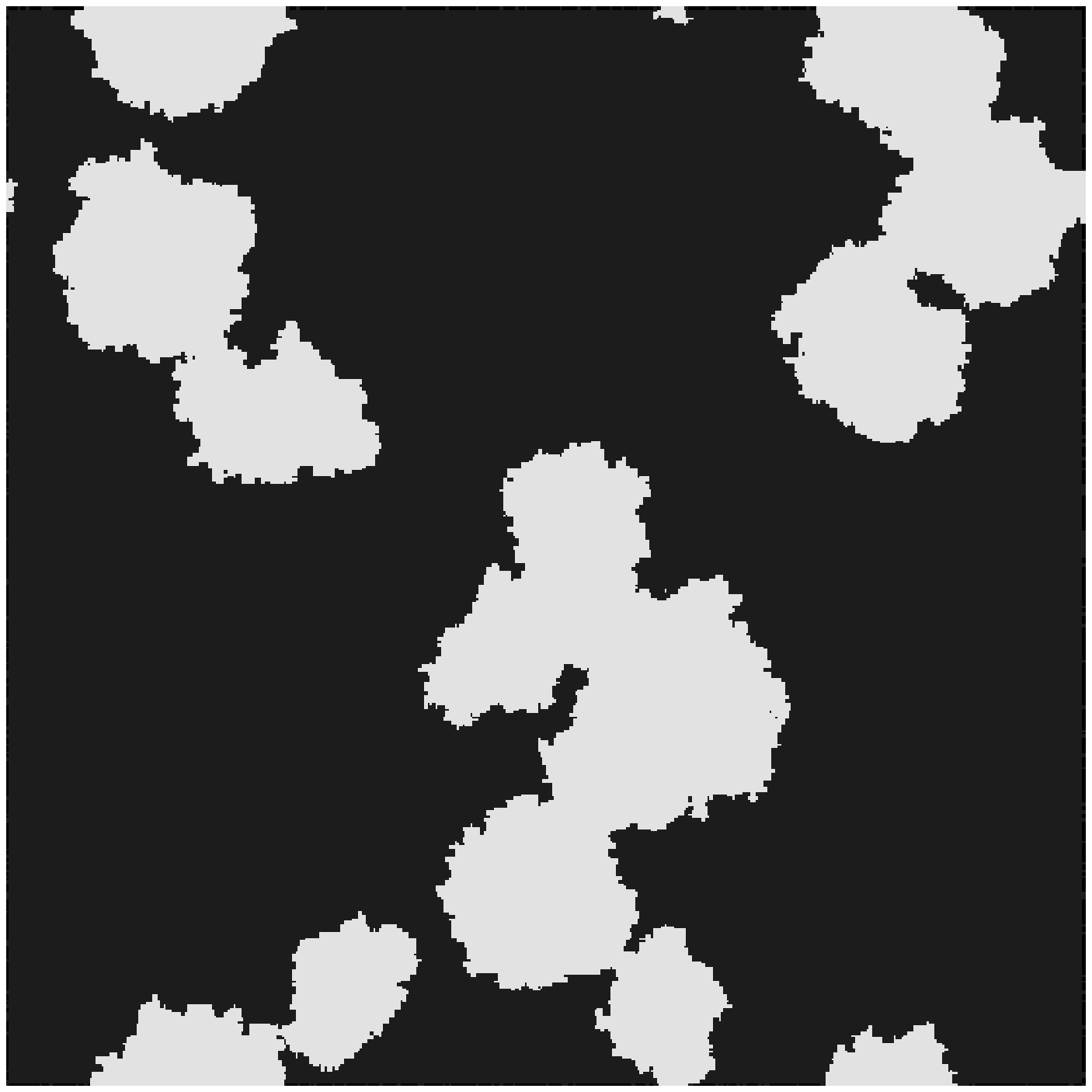}\\
 \includegraphics[width=2.2cm]{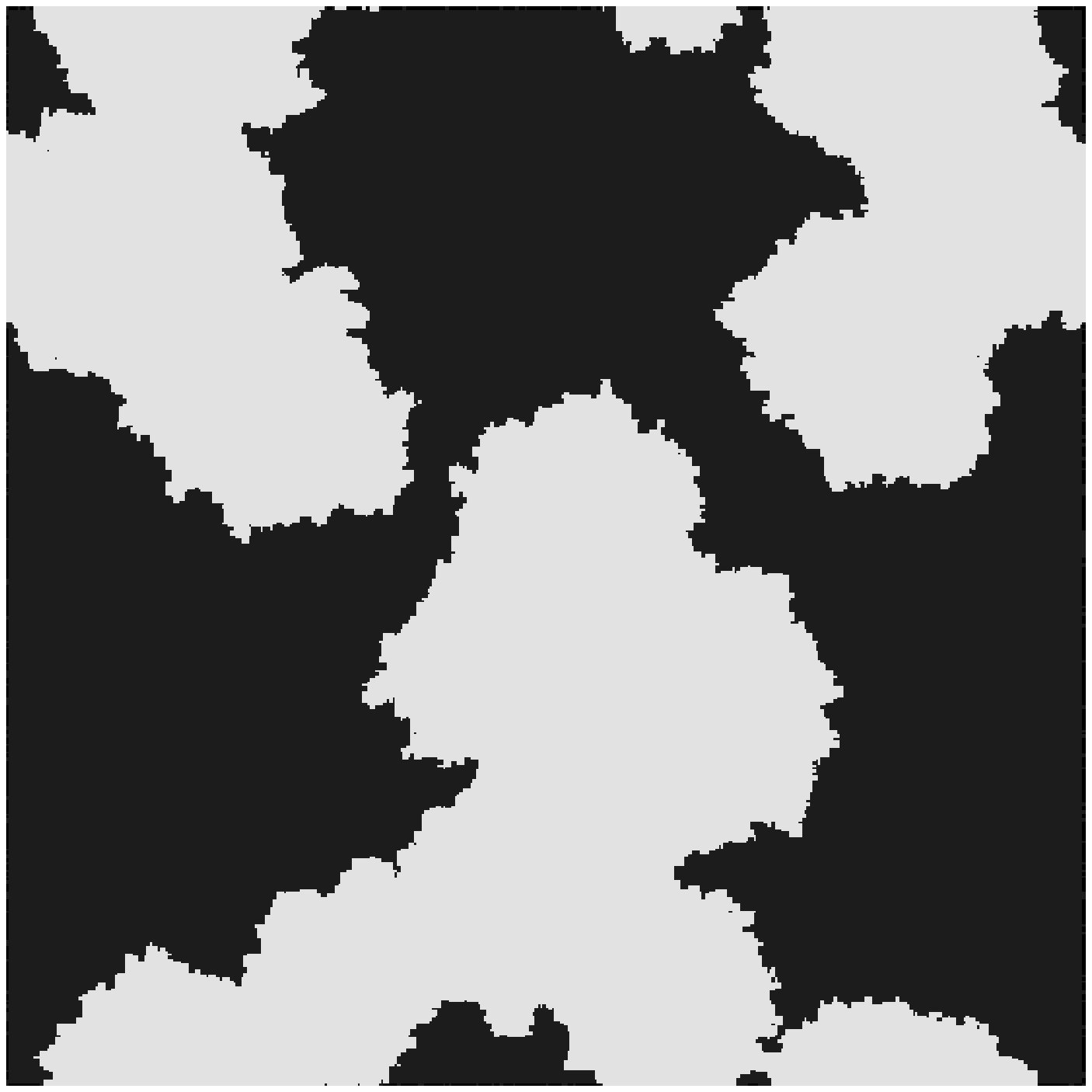}
 \includegraphics[width=2.2cm]{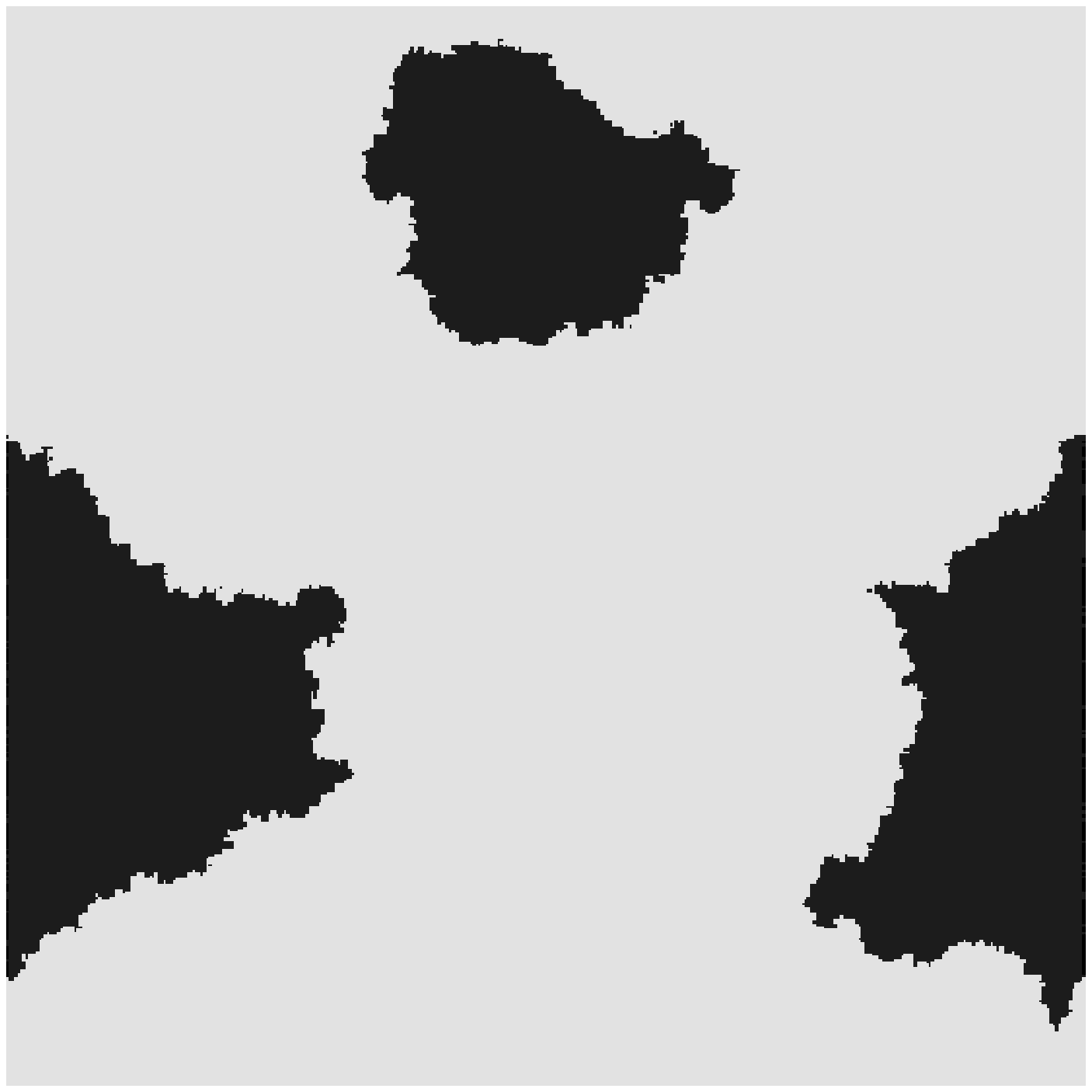}
 \includegraphics[width=2.2cm]{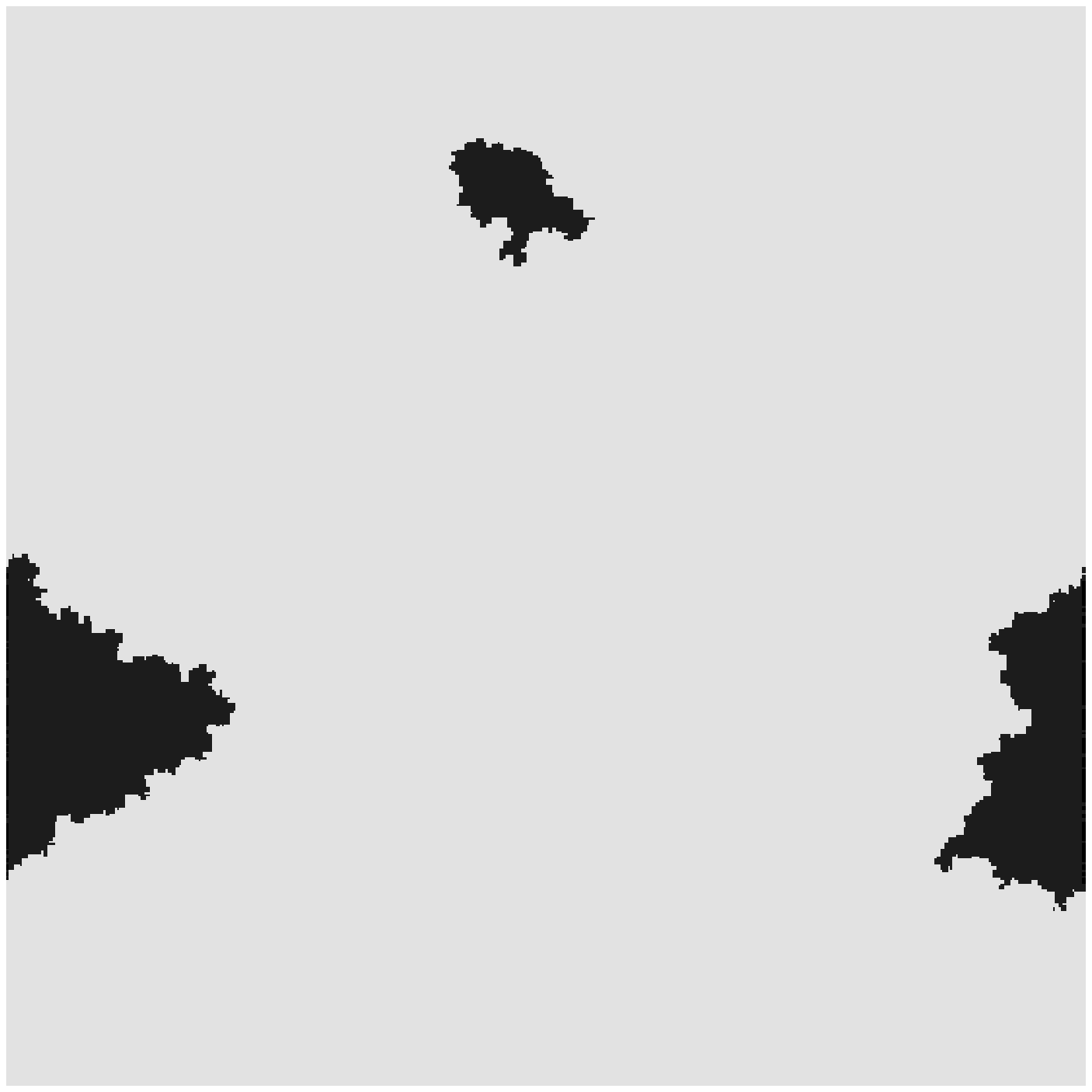}
\end{center}
\caption{\label{snap-2d} The dynamical evolution of a 2D local
model updated according to the asynchronous rule is shown. In the
simulation $N = 400 \times 400$, $M = 5$ and $p = 0.2$. The time
increases from left to right and from top to bottom. The snapshots
correspond, respectively to $T = 0$, $20$, $52$, $106$, $276$,
$394$, $594$ and $818$.}
\end{figure}

\begin{figure} 
\begin{center}
\includegraphics[width=6.8cm,angle=-90]{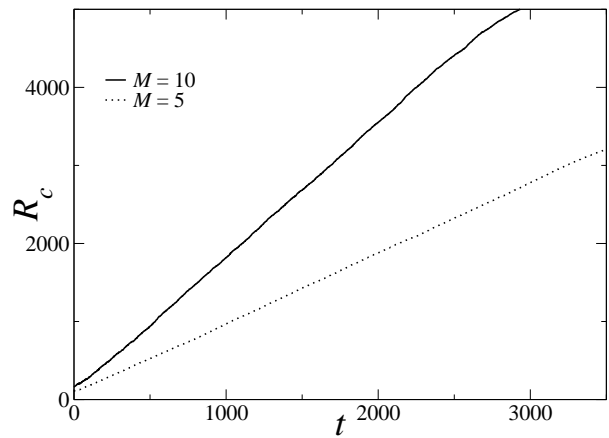} 
\end{center}
\caption{\label{rg-fig} Characteristic radius, $R_c$ averaged over 100 independent
runs as a function of time for a two-dimensional system updated according to
the synchronous update and regular tessellation.  $R_c$ is shown to grow linearly
with time for different values of $M$. The initial condition is taken as a single circular domain of minority supporters with a radius $R>R^*$.} 
\end{figure}

The 2D snapshots of the dynamical evolution, see Fig. \ref{snap-2d}, show that
as in the 1D version, domains are formed in an initial transient, and after
this, grow by interface dynamics. To characterize the growth of the
characteristic size of the domains we calculate a characteristic radius $R_c$
defined as 
\begin{equation} 
R_c^2 = \frac{\sum_{i} \delta_+(i) \vec{r_i}^2}{\sum_{i} \delta_+(i)} - \left(\frac{\sum_{i} \delta_+(i) \vec{r_i}}{\sum_{i} \delta_+(i)}  
\right)^2 . 
\end{equation} 
Where $\delta_+(i)$ is $1$ if individual $i$
supports opinion $+$, $0$ otherwise and $\vec{r_i}$ is the position in the square
lattice of $i-$th individual.  Figure \ref{rg-fig} shows that this
characteristic linear dimension grows linearly with time as in the 1D model. 

\begin{figure} \begin{center}
\includegraphics[width=6.8cm,angle=-90]{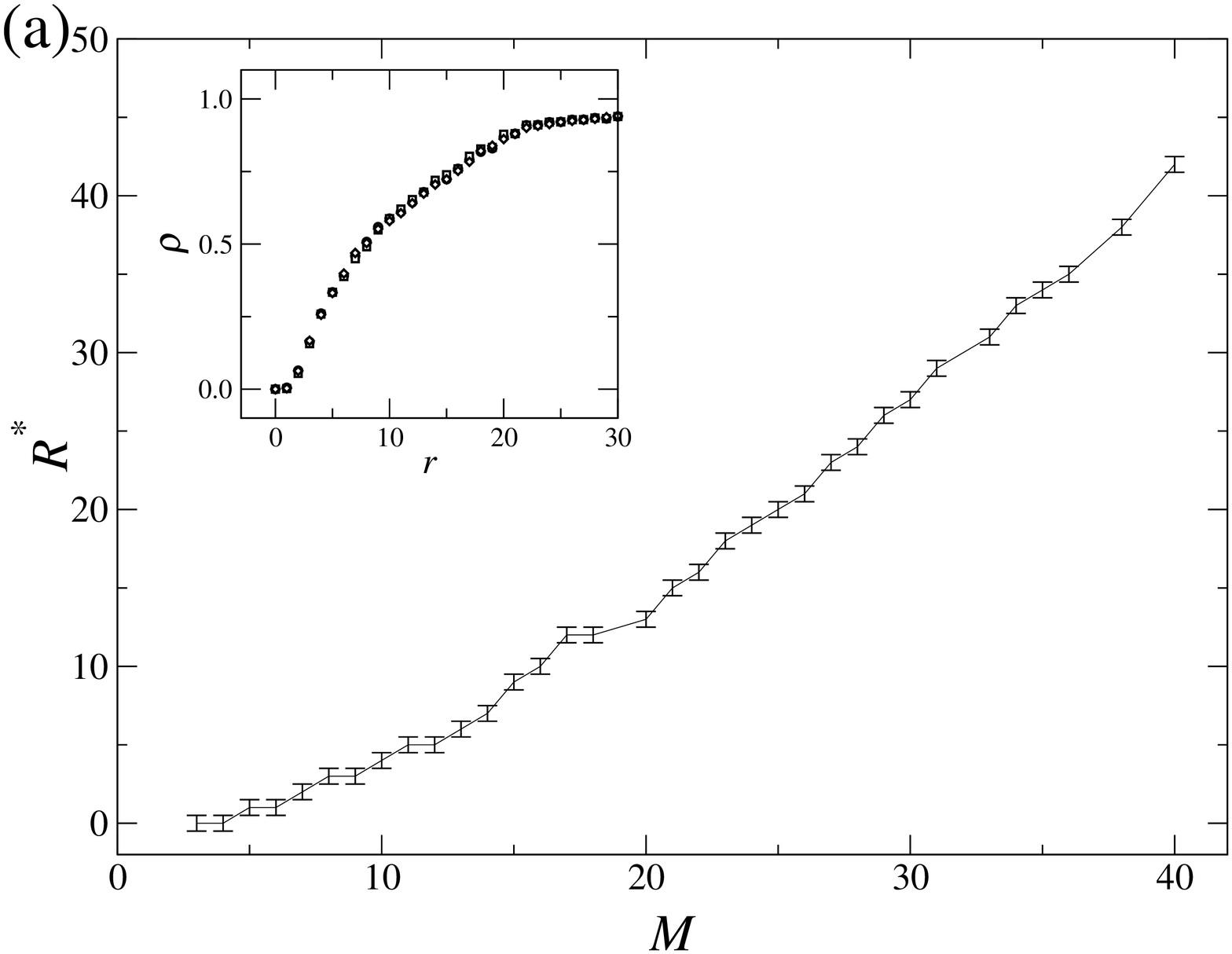}\\
\includegraphics[width=6.8cm,angle=-90]{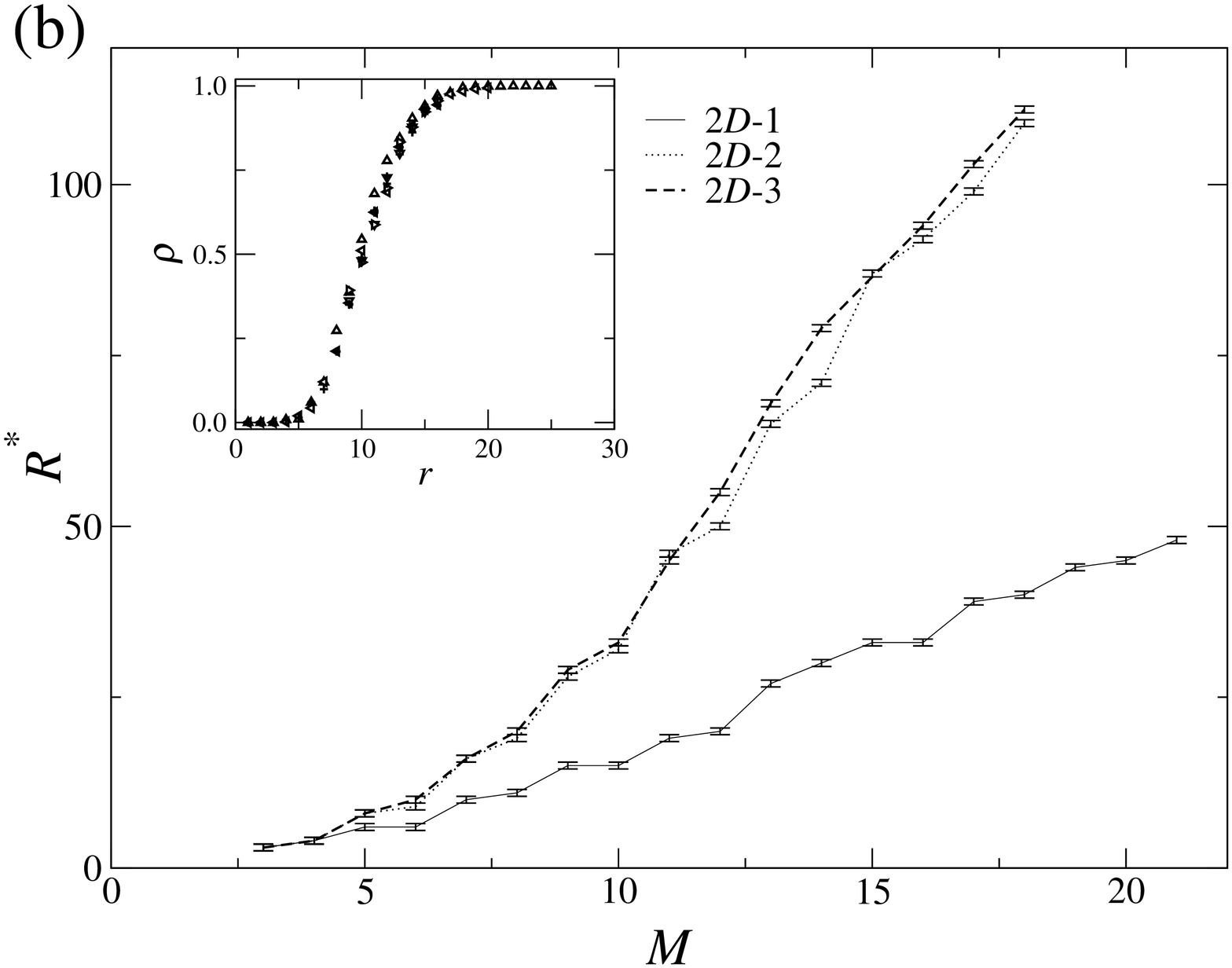}  
\end{center}
\caption{\label{rc-fig} Critical radius vs $M$, for the different versions of
the neighborhood systems: (a) 1D model. (b) 2D-1, 2D-2, 2D-3 as defined in
Fig. \ref{alpha}. The insets show the probability $\rho(r)$ that an initial
domain of radius $r$ grows; (a) corresponds to $M=15$ and (b) is for the 2D-1
model for $M=7$, both for different system sizes.}  
\end{figure}

The growth of domains in 1D and 2D occurs for those domains whose initial
characteristic size is larger than a critical one $R^*$. When the initial size
is larger than $R^*$, domains grow until they take over the whole system, while
smaller domains shrink toward extinction. The quantitative calculation of the
critical radius starts by placing a unique circular island of radius $r$ of
minority people surrounded by the majority and watching it grow or shrink. This
initial condition is quite different of that of a random initial distribution.
In order to avoid finite size effects, one has to be careful to choose a system
size $N$ such that the initial density $p = \pi r^2 / N$ is lower than the
critical one, $p_c$. In the inset of Fig. \ref{rc-fig} we show that $\rho(r)$,
the probability that an initial circle of radius $r$ grows, is indeed
independent of system size. We define the critical radius as the  value of $r$
such that $\rho(r) = 0.5$.  Figure \ref{rc-fig} shows that the value of the
critical radius increases linearly with $M$, both in 1D and in 2D models.

\section{General conclusions}

We have revisited Galam's model \cite{Gal01,Gal02} of minority
opinion spreading and introduced related neighborhood models that
incorporate spatial local effects in the interactions. These
models share basic characteristics with the {\sl bounded
neighborhood} and {\sl spatial proximity} models of Schelling
\cite{Schelling}. In both cases we have considered in detail the
role of system size in the properties of the system. For the
original nonlocal version \cite{Gal01,Gal02}, we have found that
the transition from initial minority final dominance to initial
minority disappearance is smeared out in a region of size
$N^{-1/2}$, while the time it takes to reach complete consensus
increases as $\ln N$, as in Stauffer's percolation model
\cite{stauffer11}.

In our local neighborhood models, we have considered 1D and 2D
lattices with regular and locally grown tessellations, both with
synchronous and asynchronous updates. All these local versions
behave qualitatively in the same way. The most important finding
is that the threshold value for the initial minority concentration
$p_c$ decreases as $N^{-\alpha}$, such that the transition from
initial minority spreading to majority dominance disappears in the
thermodynamic limit $N\to \infty$. The neighborhood models are, in
this sense, more efficient to spread an initially minority
opinion. However, while a nonlocal model is very fast in spreading
a rumor, the corresponding relaxation times to reach consensus in
the neighborhood models are much larger since they turn out to increase
with a power of the system size $N$.

We have also shown that the fact that $lim_{N \to \infty} p_c =0$ is due to the
existence of a critical size for a spatial domain of minority supporters. For
large enough systems there is always an over-critical domain that spreads and
occupies the whole system, with a characteristic average radius growing
linearly with time. This critical size domain has some analogies with the
critical nucleus of nucleation theory \cite{Gunton83}, but it has different
characteristics. In classical nucleation the existence of a critical radius is
due to the competition between surface tension and different bulk energy
between the two possible homogeneous states. The concept of critical nucleus in
this context is not meaningful for one-dimensional systems for which no surface
tension exists. An over-critical droplet appears as a rare fluctuation in the
bulk of a metastable state and it then grows deterministically. Noisy
perturbations in the growth dynamics are generally a second order effect. In
our case over-critical domains appear in the random initial condition. The
critical size is here an average concept resulting from the competition of the
bias favoring the minority opinion in case of a tie (analog of bulk energy
difference) and a stochastic dynamics that might lead to the disappearance of
the minority domain, surface tension being a second order effect. Critical size
means here equal probability for the domain to spread or to collapse.

The existence of this critical size clearly shows an important difference
between typical statistical physics problems and \textit{sociophysical} ones.
In the former case, one is mostly concerned with the thermodynamic limit of
large systems, while these findings emphasize the important role of system size
in the latter case.

{\bf Acknowledgments} We thank P.Colet, V. M. Egu\'{\i}luz and E.
Hern\'andez--Garc{\'\i}a for fruitful discussions. We acknowledge financial
support from MCYT (Spain) and FEDER through the projects BFM2001-0341-C02-01
and BFM2000-1108. HSW acknowledges partial support from ANPCyT, Argentine, and
thanks the MECyD, Spain, for an award within the {\it Sabbatical Program for
Visiting Professors}, and to the Universitat de les Illes Balears for the kind
hospitality extended to him.

\end{document}